\documentclass[
  journal=largetwo,
  manuscript=article-type,
  year=2024,
  volume=37,
  usenatbib,
]{cup-journal}
\usepackage{aas-macros}

\usepackage{mathrsfs}
\usepackage{amssymb}
\usepackage{comment}
\usepackage{subcaption}
\usepackage{multirow}
\usepackage{enumitem}
\usepackage{threeparttable}
\usepackage{tabularray} 
\UseTblrLibrary{booktabs}

\AtEveryBibitem{\clearfield{doi}}
\AtEveryBibitem{\clearfield{archivePrefix}}
\AtEveryBibitem{\clearfield{month}}
\AtEveryBibitem{\clearfield{eprint}}
\AtEveryBibitem{\clearfield{url}}
\AtEveryBibitem{\ifentrytype{article}{\clearfield{title}}{}}
\AtEveryBibitem{\clearfield{issn}}

\usepackage{graphicx}	
\usepackage{amsmath}	
\usepackage{scalerel,tikz}
\usetikzlibrary{svg.path}
\definecolor{orcidlogocol}{HTML}{A6CE39}
\tikzset{orcidlogo/.pic={
		\fill[orcidlogocol] svg{M256,128c0,70.7-57.3,128-128,128C57.3,256,0,198.7,0,128C0,57.3,57.3,0,128,0C198.7,0,256,57.3,256,128z};
		\fill[white] svg{M86.3,186.2H70.9V79.1h15.4v48.4V186.2z}
		svg{M108.9,79.1h41.6c39.6,0,57,28.3,57,53.6c0,27.5-21.5,53.6-56.8,53.6h-41.8V79.1z M124.3,172.4h24.5c34.9,0,42.9-26.5,42.9-39.7c0-21.5-13.7-39.7-43.7-39.7h-23.7V172.4z}
		svg{M88.7,56.8c0,5.5-4.5,10.1-10.1,10.1c-5.6,0-10.1-4.6-10.1-10.1c0-5.6,4.5-10.1,10.1-10.1C84.2,46.7,88.7,51.3,88.7,56.8z};
}}
\newcommand\orcidicon[1]{\href{https://orcid.org/#1}{\mbox{\scalerel*{
				\begin{tikzpicture}[yscale=-1,transform shape]
					\pic{orcidlogo};
				\end{tikzpicture}
			}{|}}}}

\usepackage{hyperref}
\usepackage[nopatch]{microtype}
\usepackage{booktabs}

\newcommand{\meraxes}{\ifmmode\mathrm{\textsc{Meraxes}}\else{}{\textsc{Meraxes} }\fi}

\newcommand{\Msun}{\ifmmode\mathrm{M_\odot}\else{}$\rm M_\odot$\fi}

\newcommand{\cmfast}{{\tt 21cmFAST}}

\newcommand{\xH}{\ifmmode{x}_{\rm HI}\else{}${x}_{\rm HI}$\fi}

\newcommand{\atomicseven}{{\tt Evolving\_f_{esc}}}
\newcommand{\atomicsix}{{\tt Constant\_f_{esc}}}

\defcitealias{Qin2021MNRAS.506.2390Q}{Q21}

\title{{\color{black}Percent-level} timing of reionization: self-consistent, implicit-likelihood inference from XQR-30+ Ly$\alpha$ forest data}

\author{Yuxiang Qin~\orcidicon{0000-0002-4314-1810}}
\affiliation{Research School of Astronomy and Astrophysics, Australian National University, Canberra, ACT 2611, Australia}
\alsoaffiliation{ARC Centre of Excellence for All Sky Astrophysics in 3 Dimensions (ASTRO 3D), Australia}
\email[Y. Qin]{Yuxiang.L.Qin@gmail.com}

\author{Andrei Mesinger~\orcidicon{0000-0003-3374-1772}}
\affiliation{Scuola Normale Superiore, Piazza dei Cavalieri 7, 56125 Pisa, Italy}

\author{David Prelogović~\orcidicon{0000-0002-1319-5447}}
\affiliation{Scuola Normale Superiore, Piazza dei Cavalieri 7, 56125 Pisa, Italy}
\alsoaffiliation{Scuola Internazionale Superiore di Studi Avanzati (SISSA), Via Bonomea 265, 34136 Trieste, Italy}

\author{George Becker~\orcidicon{0000-0003-2344-263X}}
\affiliation{Department of Physics \& Astronomy, University of California, Riverside, CA 92521, USA}

\author{Manuela Bischetti~\orcidicon{0000-0002-4314-021X}}
\affiliation{Dipartimento di Fisica, Universit\'a di Trieste, Sezione di Astronomia, Via G.B. Tiepolo 11, I-34131 Trieste, Italy}
\alsoaffiliation{INAF - Osservatorio Astronomico di Trieste, Via G. B. Tiepolo 11, I–34131 Trieste, Italy}

\author{Sarah E. I. Bosman~\orcidicon{0000-0001-8582-7012}}
\affiliation{Institute for Theoretical Physics, Heidelberg University, Philosophenweg 12, D–69120, Heidelberg, Germany}
\alsoaffiliation{Max-Planck-Institut f\"{u}r Astronomie, K\"{o}nigstuhl 17, D-69117 Heidelberg, Germany}

\author{Frederick B. Davies~\orcidicon{0000-0003-0821-3644}}
\affiliation{Max-Planck-Institut f\"{u}r Astronomie, K\"{o}nigstuhl 17, D-69117 Heidelberg, Germany}

\author{Valentina D'Odorico~\orcidicon{0000-0003-3693-3091}}
\affiliation{INAF - Osservatorio Astronomico di Trieste, Via G. B. Tiepolo 11, I–34143 Trieste, Italy}
\alsoaffiliation{Scuola Normale Superiore, Piazza dei Cavalieri 7, 56125 Pisa, Italy}

\author{Prakash Gaikwad~\orcidicon{0000-0002-2423-7905}}
\affiliation{Max-Planck-Institut f\"{u}r Astronomie, K\"{o}nigstuhl 17, D-69117 Heidelberg, Germany}

\author{Martin G. Haehnelt~\orcidicon{0000-0001-8443-2393}}
\affiliation{Kavli Institute for Cosmology and Institute of Astronomy, Madingley Road, Cambridge, CB3 0HA, UK}

\author{Laura Keating~\orcidicon{0000-0001-5211-1958}}
\affiliation{Institute for Astronomy, University of Edinburgh, Blackford Hill, Edinburgh, EH9 3HJ, UK}

\author{Samuel Lai~\orcidicon{0000-0001-9372-4611}}
\affiliation{Commonwealth Scientific and Industrial Research Organisation (CSIRO), Space \& Astronomy, P. O. Box 1130, Bentley, WA 6102, Australia}

\author{Emma Ryan-Weber~\orcidicon{0000-0002-5360-8103}}
\affiliation{Centre for Astrophysics and Supercomputing, Swinburne University of Technology, Hawthorn, Victoria 3122, Australia}
\alsoaffiliation{ARC Centre of Excellence for All Sky Astrophysics in 3 Dimensions (ASTRO 3D), Australia}

\author{Sindhu Satyavolu~\orcidicon{0000-0001-5818-6838}}
\affiliation{Tata Institute of Fundamental Research, Homi Bhabha Road, Mumbai 400005, India}
\alsoaffiliation{Institut de Física d’Altes Energies (IFAE), The Barcelona Institute of Science and Technology, Campus UAB, 08193 Bellaterra Barcelona, Spain}

\author{Fabian Walter~\orcidicon{0000-0003-4793-7880}}
\affiliation{Max-Planck-Institut f\"{u}r Astronomie, K\"{o}nigstuhl 17, D-69117 Heidelberg, Germany}
\alsoaffiliation{National Radio Astronomy Observatory, Pete V. Domenici Array Science Center, P.O. Box O, Socorro, NM 87801, USA}

\author{Yongda Zhu~\orcidicon{0000-0003-3307-7525}}
\affiliation{Steward Observatory, University of Arizona, 933 North Cherry Avenue, Tucson, AZ 85721, USA}


\begin{document}

\begin{abstract}
The Lyman alpha (Ly$\alpha$) forest in the spectra of $z>5$ quasars provides a powerful probe of the late stages of the Epoch of Reionization (EoR).  With the recent advent of exquisite datasets such as XQR-30, many models have struggled to reproduce the observed large-scale fluctuations in the Ly$\alpha$ opacity.
Here we introduce a Bayesian analysis framework that forward-models large-scale lightcones of intergalactic medium (IGM) properties, and accounts for unresolved sub-structure in the Ly$\alpha$ opacity by calibrating to higher-resolution hydrodynamic simulations.
Our models directly connect physically-intuitive galaxy properties with the corresponding IGM evolution, without having to tune ``effective'' parameters or calibrate out the mean transmission.  
The forest data, in combination with UV luminosity functions and the CMB optical depth, are able to constrain global IGM properties at percent level precision in our fiducial model.
Unlike many other works, we recover the forest observations without
evoking a rapid drop in the ionizing emissivity from $z\sim7$ to 5.5, {\color{black}which we attribute to our sub-grid model for recombinations}.  
In this fiducial model, reionization ends at $z=5.44\pm0.02$ and the EoR mid-point is at $z=7.7\pm0.1$.  The ionizing escape fraction increases towards faint galaxies, showing a mild redshift evolution at fixed UV magnitude, $M_{\rm UV}$.  Half of the ionizing photons are provided by galaxies fainter than $M_{\rm UV} \sim -12$, well below direct detection limits of optical/NIR instruments including ${\it JWST}$.
We also show results from an alternative galaxy model that does not allow for a redshift evolution in the ionizing escape fraction.  Despite being decisively disfavored by the Bayesian evidence, the posterior of this model is in qualitative agreement with that from our fiducial model.
We caution however that our conclusions regarding the early stages of the EoR and which sources reionized the Universe are more model-dependent.
\end{abstract}

\section{Introduction}

The Epoch of Reionization (EoR) is a fundamental milestone in the evolution of our Universe. Its timing and spatial fluctuations encode invaluable information about the intergalactic medium (IGM) and the first galaxies.
Recent years have witnessed a dramatic increase in the number and quality of observations probing the EoR, including upper limits on the cosmic 21-cm power spectrum \citep{Mertens2020MNRAS.493.1662M,Trott2020MNRAS.493.4711T,HERA2023ApJ...945..124H}, the polarization anisotropy of the cosmic microwave background (CMB; \citealt{Planck2020A&A...641A...6P,Reichardt2021ApJ...908..199R}), and the IGM Lyman-$\alpha$ (Ly$\alpha$) damping-wing absorption seen in spectra of high-redshift quasars \citep{Banados2018Natur.553..473B,Wang2020ApJ...896...23W} and star-forming galaxies  \citep{Pentericci2018A&A...619A.147P,Umeda2024ApJ...971..124U,Heintz2024arXiv240402211H}.

Arguably the most mature of EoR datasets is the Ly$\alpha$ forest. More than two decades of observational efforts have provided over 70 high-quality quasar spectra at $z>5.5$ \citep{Fan2002AJ....123.1247F,Fan2006AJ....131.1203F,Fan2006AJ....132..117F,Willott2007,Becker2015MNRAS.447.3402B,Wu2015Natur.518..512W,Banados2016ApJS..227...11B,Jiang2016ApJ...833..222J,Eilers2018ApJ...864...53E,Yang2020ApJ...897L..14Y,DOdorico2023MNRAS.523.1399D}. These data provide unparalleled statistics over large volumes of the IGM. As such, the Ly$\alpha$ forest is one of the few EoR probes that is not sensitive to the biased environments proximate to the ionizing sources.

{\color{black}The high quality and quantity of Lya forest data} provide an invaluable stress test on our understanding of the EoR, as they are quite sensitive to missing {\color{black}components} in our theoretical and systematic models. For instance, the observed large-scale fluctuations in the Ly$\alpha$ optical depth cannot be reproduced by the simplest, uniform ultraviolet background (UVB) models at $z>5.2$ \citep{Becker2015MNRAS.447.3402B,Bosman2022MNRAS.514...55B}.  Various theoretical models have attempted to reproduce the observations by increasing fluctuations in the IGM temperature, mean free path (MFP) of ionizing photons, ionizing emissivity, and/or including an ongoing, patchy reionization \citep{DAloisio2015ApJ...813L..38D,Davies2016MNRAS.460.1328D,DAloisio2017MNRAS.468.4691D,DAloisio2018MNRAS.473..560D,Chardin2017MNRAS.465.3429C,Kulkarni2019MNRAS.485L..24K,Keating2020MNRAS.491.1736K,Meiksin2020MNRAS.491.4884M,Nasir2020MNRAS.494.3080N,Asthana2024arXiv240406548A}. However, moving beyond ``\textit{this particular model is (in)consistent with the data}''  to ``\textit{this is the distribution of IGM and galaxy properties inferred from the data}'' is considerably challenging, and can only be achieved in a physically-motivated, efficient Bayesian inference framework.

Previous work that reproduced the data relied heavily on \textit{effective} (i.e. not physically \-interpretable) parameters and/or ad-hoc assumptions that ignore or fine-tune the redshift evolution of the mean transmission flux. For example, several studies found that in order to reproduce the forest data, the UV ionizing emissivity in their simulations has to be tuned to drop rapidly towards the end of the EoR, with up to a factor of 2 decrement over just $\Delta z\sim0.5$ (${\sim}100$ Myr at these redshifts; e.g., \citealt{Kulkarni2019MNRAS.485L..24K,Ocvirk2021MNRAS.507.6108O}; Fig. \ref{fig:post_emissivity}). Such short time-scales for the UVB evolution are difficult to justify physically (e.g., \citealt{Sobacchi2013MNRAS.432.3340S}) or to reconcile with the observed gradual evolution of the cosmic star formation rate (SFR) density from  bright galaxies \citep{Bouwens2015ApJ...803...34B,Oesch2018ApJ...855..105O}. Indeed subsequent analysis pointed to unresolved substructure in the simulations as a possible explanation (e.g., see section 5.4 in \citealt{Qin2021MNRAS.506.2390Q}, and the recent analysis in \citealt{Cain2024MNRAS.531.1951C}).
Alternatively, simulations that tune the ionizing MFP without modelling the time evolution of HII regions and/or adopt effective parameters for inhomogeneous recombinations are also difficult to interpret as they only provide a somewhat opaque proxy for cosmic reionization (e.g., \citealt{Choudhury2020arXiv200308958C,Gaikwad2023MNRAS.525.4093G,Davies2024ApJ...965..134D}).

Ideally, one should use a self-consistent model in which the redshift evolution of the patchy reionization is simulated directly from the galaxies that drive it.  This would allow us to set well-motivated priors on {\it physical} parameters that can be constrained by complementary galaxy observations (e.g., \citealt{Park2019MNRAS.484..933P,Mutch2024MNRAS.527.7924M}). Anchoring the EoR models on galaxies also allows us to constrain earlier epochs where we have no forest measurements, since structure evolution (i.e., the halo mass function) is comparably well understood (e.g., \citealt{Sheth2001}) and we have complementary  observations of UV luminosity functions (LFs) that constrain how halos are populated with galaxies at these high redshifts.

However, such self-consistent modelling of the EoR is inherently extremely challenging, due to the enormous dynamic range of relevant scales. Fluctuations in the Ly$\alpha$ forest are correlated on scales larger than ${\sim}100$ cMpc (e.g., \citealt{Becker2021MNRAS.508.1853B,Zhu2021ApJ...923..223Z}), while galaxies and IGM clumps are on sub-kpc scales (e.g., \citealt{Schaye2001ApJ...559..507S, Emberson2013ApJ...763..146E, Park2016ApJ...831...86P, DAloisio2020ApJ...898..149D}). As a result, current simulations must rely on sub-grid prescriptions that have to be calibrated against observations or other more detailed, higher resolution simulations.

Here, we present an updated Bayesian inference framework for the high-redshift Ly$\alpha$ forest that is arguably free from ``effective'' parameters. We sample physically-intuitive galaxy scaling relations to compute large-scale lightcones of the Ly$\alpha$ opacity using {\cmfast} \citep{Mesinger2011MNRAS.411..955M,Murray2020JOSS....5.2582M}. This self-consistently connects galaxy properties to the state of the IGM that is shaped by their radiation fields.
We account for missing small-scale structure by calibrating to the Sherwood suite of high-resolution hydrodynamic simulations \citep{Bolton2017MNRAS.464..897B}. This calibration
allows us to eliminate the poorly-motivated hyperparameters we previously used to account for missing systematics and/or physics (\citealt{Qin2021MNRAS.506.2390Q}, hereafter \citetalias{Qin2021MNRAS.506.2390Q}). 
For each astrophysical parameter combination, we forward model the forest transmission, comparing against the observations \citep{Bosman2022MNRAS.514...55B} using an {\it implicit} likelihood.  We present the resulting joint constraints on reionization and galaxy properties, implied by the combined data from the Ly$\alpha$ forest, UV LFs, and CMB optical depth.

This paper is organized as follows. We summarize the extended XQR-30 Ly$\alpha$ forest data in Section \ref{sec:obs}, and introduce our Bayesian framework for forward-modelling Ly$\alpha$ forests in Section \ref{sec:model}. After summarizing the complementary  observations and free parameters used in this work in Sections \ref{subsec:likelihood_other} and \ref{sec:inference}, we present results in Section \ref{sec:result} including the recovered properties of the IGM and those of the underlying galaxies. We then discuss the implication to our understanding of reionization in Section \ref{sec:model_dependence}, before concluding in Section \ref{sec:conclusion}. In this work, we adopt cosmological parameters from Planck ($\Omega_{\mathrm{m}}, \Omega_{\mathrm{b}}, \Omega_{\mathrm{\Lambda}}, h, \sigma_8, n_\mathrm{s} $ = 0.312, 0.0490, 0.688, 0.675, 0.815, 0.968; \citealt{Planck2016A&A...594A..13P}). {\color{black}Distance units are comoving unless otherwise specified.}

\section{The Ly$\alpha$ opacity distributions from XQR-30+}
\label{sec:obs}

The ultimate XSHOOTER legacy survey of quasars at $z\sim5.8$--6.6 (XQR-30) is a ${\sim}250$-hour programme using the Very Large Telescope (VLT) at the European Southern Observatory (ESO; \citealt{DOdorico2023MNRAS.523.1399D}). While XQR-30 contains 30 high-quality quasar spectra, \citet{Bosman2022MNRAS.514...55B} assembled 67 sightlines at these redshifts by combining XQR-30 with archival spectra. We refer to this extended dataset as XQR-30+.

\begin{figure*}
\centering
    \includegraphics[width=\textwidth]{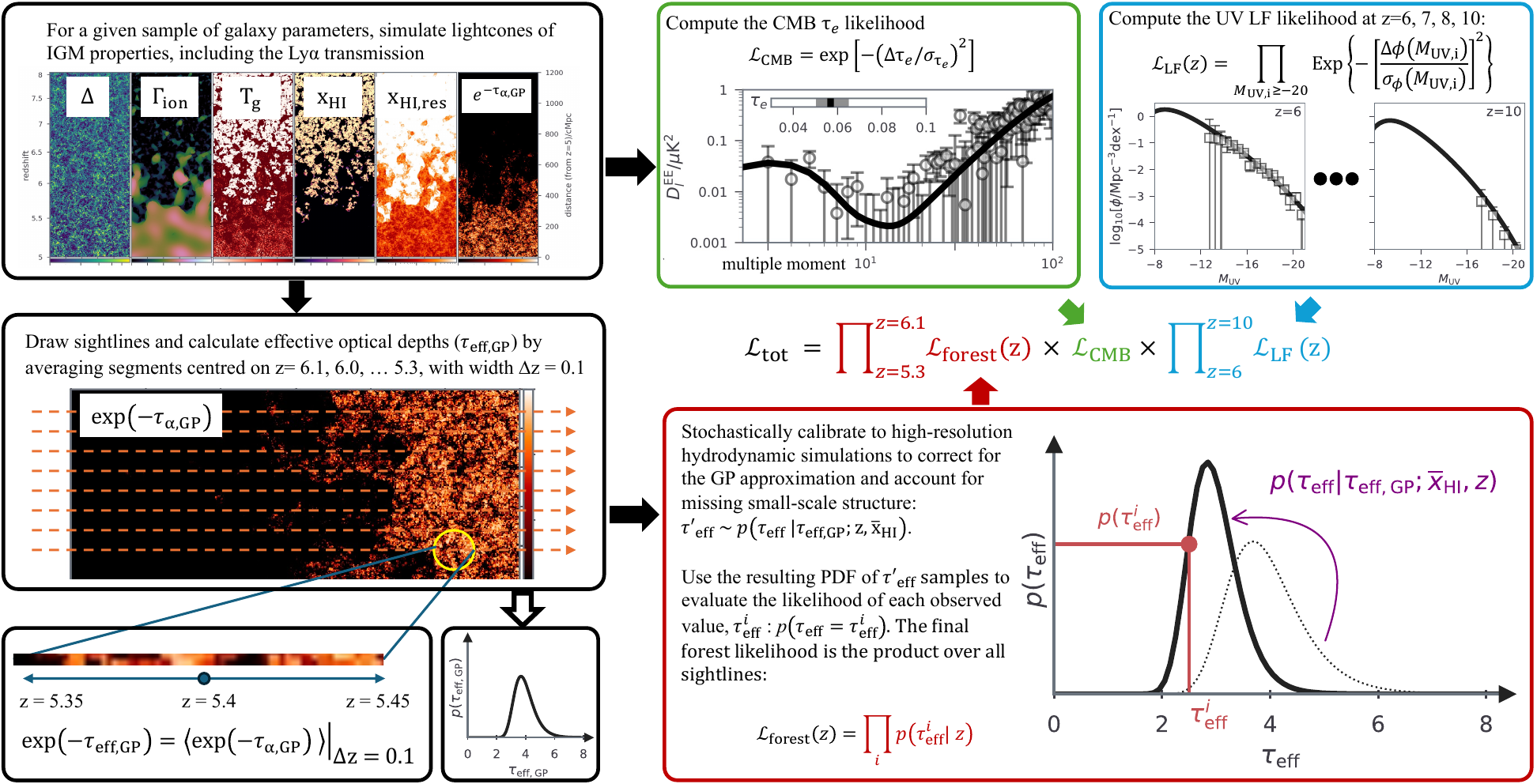}\vspace*{-2.8mm}
    \caption{A flow chart showing the steps involved in computing the likelihood for a single sample of astrophysical parameters.  See text for more details.\label{fig:reference}}
\end{figure*}

The Ly$\alpha$ transmission in these spectra was quantified by the commonly-used ``effective optical depth'', $\tau_{\rm eff} \equiv - \ln \langle\mathcal{F}_\alpha\rangle_{\Delta z=0.1}$.  Here $\mathcal{F}_\alpha(\lambda)$ 
is the continuum-normalized flux in the Ly$\alpha$ forest, which is averaged over segments of width $\Delta z=0.1$ (roughly corresponding to $\sim$ 40 cMpc at these redshifts). Non-detections {\color{black}(2$\sigma$)} were assigned lower limits on $\tau_{\rm eff}$ corresponding to twice the mean flux noise in the corresponding segment.   The full XQR-30+ sample has at least ${\sim}10$ estimates of $\tau_{\rm eff}$ in each redshift bin spanning $z=5.1$, 5.2, ..., 6.1. We show the cumulative distribution functions (CDFs) of these $\tau_{\rm eff}$ estimates in Fig. \ref{fig:post_CDF}, where we also compare them to our fiducial posterior.  For more details on how the observations were processed, see \citet{Bosman2022MNRAS.514...55B}.

\section{Forward modelling}\label{sec:model}
We use the public simulation code, \cmfast\ \footnote{\url{https://github.com/21cmfast/21cmFAST}}\citep{Mesinger2007ApJ...669..663M,Mesinger2011MNRAS.411..955M,Murray2020JOSS....5.2582M}, to compute 3D lightcones of the Ly$\alpha$ IGM opacity. 
 A single forward model and the corresponding likelihood evaluation are summarized in the flow chart of Fig. \ref{fig:reference} and consist of the following steps:
\begin{enumerate}[labelwidth=!,itemindent=24pt,labelindent=0pt, leftmargin=0em, itemsep=3pt, parsep=0pt, topsep = 3pt]
    \item Simulate large-scale 3D lightcones of the IGM density ($\Delta\equiv\rho/\overline{\rho}$), neutral fraction due to inhomogeneous reionization ($x_{\rm HI}$), photo-ionization rate ($\Gamma_{\rm ion}$), IGM temperature ($T_{\rm g}$), residual neutral fraction inside the ionized IGM ($x_{\rm HI,res}$) and corresponding Ly$\alpha$ opacity (top left panel of Fig. \ref{fig:reference});
    \item Construct mock quasar sightlines and compute the effective optical depth by binning the sightlines over the same redshift intervals as the XQR-30+ observation (lower left panels of Fig. \ref{fig:reference});
    \item Account for missing small scales by calibrating these effective optical depths against high-resolution hydrodynamic simulations.  Use the resulting probability density function (PDF) of calibrated  $\tau_{\rm eff}$ to evaluate the likelihood of the observed values (lower right panel of Fig. \ref{fig:reference});
    \item Multiply this forest likelihood with the corresponding UV LF and CMB likelihoods in order to obtain the total likelihood of this parameter sample (upper right panels in Fig. \ref{fig:reference}; c.f. Section \ref{subsec:likelihood_other}).
\end{enumerate}

We discuss this procedure in detail below, emphasizing the improvements over our previous analysis in \citetalias{Qin2021MNRAS.506.2390Q}.

\subsection{Galaxy models}\label{subsec:galaxies}

Our galaxy models are based on the semi-empirical parametrization in \cite{Park2019MNRAS.484..933P}.  We assume power laws relating the fraction of galactic baryons in stars ($f_*$) and the UV ionizing escape fraction ($f_{\rm esc}$) to the host halo mass ($M_{\rm vir}$):
\begin{equation}\label{eq:fstar}
	f_* = \min\left[1, f_{*,10}\left(\frac{M_{\rm vir}}{10^{10}\Msun}\right)^{\rm \alpha_*}\right] 
\end{equation}
and
\begin{equation}\label{eq:fesc}
	f_{\rm esc} = \min\left[1, f_{\rm esc,10}\left(\frac{M_{\rm vir}}{10^{10}\Msun}\right)^{\alpha_{\rm esc}}\left(\frac{1+z}{8}\right)^{\beta_{\rm esc}}\right],
\end{equation}
where $f_{*,10}$, $\alpha_*$, $f_{\rm esc,10}$, ${\alpha_{\rm esc}}$ and ${\beta_{\rm esc}}$ are free parameters.  Compared to \cite{Park2019MNRAS.484..933P} and our previous analysis in \citetalias{Qin2021MNRAS.506.2390Q}, here we allow for an additional redshift dependence of $f_{\rm esc}$ at a given halo mass through the parameter $\beta_{\rm esc}$ (e.g., \citealt{Haardt2012,Kuhlen2012MNRAS.423..862K,Mutch2016}).
Note that $f_*$ and $f_{\rm esc}$ have to be in the range from zero to unity as they are fractions.

\begin{table*}
\centering
\caption{Posterior distribution ([16, 84]th percentiles) and Bayesian evidence of the galaxy models used in this work. The Bayes ratio indicates a {\it very strong} preference for the $\atomicseven$ model, according to Jeffrey's scale (e.g., \citealt{Jeffreys1939thpr.book.....J})
 }\label{tab:source_model}

\begin{threeparttable}
 \begin{tabular}{l|c|c|c|c|c|c|c|l}
 \hline
   & $\log_{10} f_{*, 10}$ & $\alpha_*$ & $\log_{10}f_{\rm esc, 10}$& $\alpha_{\rm esc}$& $\beta_{\rm esc}$ & $\tau_*$ & $\log_{10}\left(M_{\rm turn}/{\rm M}_\odot\right)$ & $\ln{\mathcal{B}}$\tnote{$\bot$}\\
 \hline
 \hline
 Prior range& [-2, -0.5] & [0, 1]&[-3, 0]& [-1, 0.5]& [-3, 3] & (0, 1] & [8, 10]& --\\
 $\atomicseven$\tnote{$\dagger$} & $-1.51\pm0.03$&$0.48\pm0.05$&$-1.52_{-0.10}^{+0.12}$&$-0.94_{-0.04}^{+0.09}$&$-1.61_{-0.21}^{+0.27}$&$0.27_{-0.01}^{+0.02}$&$8.10_{-0.07}^{+0.16}$& 1 \\
 $\atomicsix$\tnote{$\ddagger$}  & $-1.42\pm0.04$&$0.51\pm0.07$&$-1.08_{-0.06}^{+0.09}$&$-0.46_{-0.13}^{+0.09}$& fixed at 0 &$0.34\pm0.02$&$8.46_{-0.35}^{+0.51}$& -17.5 \\

\hline
\end{tabular}
\begin{tablenotes}
\item[$\bot$] Bayes ratio w.r.t. $\atomicseven$ in natural logarithmic scale.
\item[$\dagger$] Galaxies have a mass-dependent and time-evolving escape fraction.
\item[$\ddagger$] Galaxies have a mass-dependent and time-independent escape fraction.
\end{tablenotes}
\end{threeparttable}
\end{table*}

The average star formation rates (SFRs) of galaxies are obtained with SFR=$M_\ast/ [\tau_\ast H^{-1}(z)]$, where $M_\ast \equiv f_\ast M_{\rm vir} {\Omega_{\mathrm{b}}}/{\Omega_{\mathrm{m}}}$ is the stellar mass, and $\tau_\ast$ is an additional free parameter corresponding to the characteristic star formation time-scale in units of the Hubble time, $H^{-1}(z)$, which scales as the halo dynamical time during matter domination. {\color{black} In this work, we adopt the conversion factor, $L_{\rm UV}/{\rm SFR} = 8.7\times 10^{27} {\rm erg}\ {\rm s}^{-1}\ {\rm Hz}^{-1}\ {\rm M}_\odot^{-1} {\rm yr}$ when calculating the UV non-ionizing luminosity.}

We also assume only a fraction $f_{\rm duty}\equiv\exp[ -M_{\rm turn}/M_{\rm vir}] $ of halos host star-forming galaxies. Here, $M_{\rm turn}$ characterizes the halo mass below which star formation becomes inefficient due to feedback and/or atomic cooling limits and is left as a free parameter.

Below we explore two galaxy models, differing in their treatment of the ionizing escape fraction:
\begin{enumerate}[labelwidth=!,itemindent=24pt,labelindent=0pt, leftmargin=0em, itemsep=3pt, parsep=0pt, topsep = 3pt]
\item $\atomicsix$ - the ionizing escape fraction is a function of halo mass only and is constant with redshift (fixing ${\beta_{\rm esc}}$ to zero in equation \ref{eq:fesc}). {\color{black}Note that this model does effectively allow for the population-averaged escape fraction to evolve with redshift, since $f_{\rm esc}$ depends on halo mass and the halo mass function evolves with redshift.  This sets a ``characteristic'' halo mass that drives both the timing and morphology of reionization.}
\item $\atomicseven$ - the ionizing escape fraction is a function of halo mass and evolves with redshift (treating both ${\alpha_{\rm esc}}$ and ${\beta_{\rm esc}}$ as free parameters in equation \ref{eq:fesc}). {\color{black}Note that adding an explicit redshift dependence to the escape fraction at a fixed halo mass gives the $\atomicseven$ model the flexibility to decouple the EoR/UVB morphology from the mean EoR history.}
\end{enumerate}
In this work we perform inference with both models, comparing their Bayesian evidences.  We find that the data strongly prefer $\atomicseven$, and we therefore refer to this model as ``fiducial''. We list the posterior distribution and Bayesian evidence of these two models in Table \ref{tab:source_model}.

\subsection{Large-scale IGM simulations}\label{subsec:igm_sim}

Our simulation boxes are 250 cMpc on a side.  Realizations of Gaussian initial conditions are computed at $z=300$ on a 640$^3$ grid, with the density fields evolved down to $z=5$ using second order Lagrangian perturbation theory {\color{black}(2LPT; \citealt{Scoccimarro1998MNRAS.299.1097S})} and smoothed down to a final resolution of 128$^3$.  Galaxy abundances are identified from the evolved density fields using excursion-set theory (\citealt{Mesinger2011MNRAS.411..955M}), and assigned properties {\color{black}including the stellar mass, SFR, ionizing escape fraction and duty cycle} according to the galaxy models discussed in the previous section.

Reionization is modelled with the excursion-set approach \citep{Furlanetto2004ApJ...613....1F}, accounting for inhomogeneous recombinations \citep{Sobacchi2014MNRAS.440.1662S}.  Unlike \citetalias{Qin2021MNRAS.506.2390Q}, here we include a correction for photon-conservation \citep{Park2022MNRAS.517..192P}, which further decreases the need for the nuisance hyperparameters used in our previous work. Specifically, a cell is flagged as ionized when the cumulative number of ionizing photons reaching it
exceeds its cumulative number of recombinations (accounting for unresolved substructure with the analytic framework of \citealt{Sobacchi2014MNRAS.440.1662S}). The former is computed by integrating over the local galaxy emissivity 
in spherical regions around each cell for radii $R \leq R_{\rm MFP, LLS}$. Here $R_{\rm MFP, LLS}$ corresponds to the MFP through the {\it ionized} IGM and is governed by {\color{black}damped Ly$\alpha$ systems (DLAs), Lyman limit systems (LLSs) and other unresolved systems with lower column densities \citep{Nasir2021ApJ...923..161N,Feron2024MNRAS.532.2401F}}

Before the end of the EoR, the {\it total} MFP determining the local ionizing background is set by a combination of $R_{\rm MFP, LLS}$ and the distance to the surrounding neutral IGM, $R_{\rm MFP, EoR}$, i.e. $R_{\rm MFP}^{-1} = R_{\rm MFP, LLS}^{-1} + R_{\rm MFP, EoR}^{-1}$ (e.g., \citealt{Alvarez2012ApJ...747..126A}).  The reionization topology computed with our excursion-set algorithm determines the local (inhomogeneous) $R_{\rm MFP, EoR}$ around each cell.  However, since we do not directly resolve the spatial distribution of LSSs and DLAs when these become rare/biased, we assume a homogeneous 
 value for
 $R_{\rm MFP, LLS}=66\left[\left(1+z\right)/6.3\right]^{-4.3}$ cMpc 
 at $z\leq6$ motivated by post-EoR measurements (\citealt{Worseck2014MNRAS.445.1745W}; see also \citealt{Songaila2010ApJ...721.1448S} and \citealt{Becker2021MNRAS.508.1853B}).\footnote{At $z>6$ where we do not have direct measurements, we set $R_{\rm MFP, LLS}=42$ cMpc.  The value of  $R_{\rm MFP, LLS}$ at these high redshifts is highly uncertain, depending on the heating history of the IGM \citep{Emberson2013ApJ...763..146E,Park2016ApJ...831...86P,DAloisio2020ApJ...898..149D}.
However, during reionization the MFP is dominated by the reionization topology (i.e. $R_{\rm MFP, LLS} > R_{\rm MFP, EoR}$; {\color{black}see Fig. \ref{fig:post_global}} and \citealt{Sobacchi2014MNRAS.440.1662S}).  Thus the exact value of $R_{\rm MFP, LLS}$ at $z>6$ should have a negligible impact on the EoR and the corresponding Ly$\alpha$ opacity distributions for realistic scenarios (see also \citealt{Cain2023MNRAS.522.2047C}).} {\color{black} In future work we will expand our model to additionally sample the mean and variance of $R_{\rm MFP, LLS}$, allowing us to extend our analysis to even lower redshifts.}

With the above, we compute the local ionizing background as
\begin{equation}\label{eq:gamma}
	{\Gamma}_{\rm ion} = \left(1+z\right)^2 R_{\rm MFP}\sigma_{\rm H}\frac{\alpha_{\rm UVB}}{\alpha_{\rm UVB}+\beta_{\rm H}}{{\dot{n}}_{\rm ion}},
\end{equation}
where the emissivity, $\dot{n}_{\rm ion}(R_{\rm MFP})$, is averaged over the local MFP around each cell, $\alpha_{\rm UVB}=2$ corresponds to the effective spectral index of the UVB (see \citealt{Becker2013MNRAS.436.1023B,DAloisio2019ApJ...874..154D}), $\beta_{\rm H}=2.75$ and $\sigma_{\rm H}=6.3\times10^{-18}{\rm cm}^2$ characterize the photo-ionization cross-section 
$\sigma(\nu) =\sigma_{\rm H} \allowbreak \left(\frac{\nu}{\nu_{\rm H}}\right)^{\beta_{\rm H}} $ with
$\nu_{\rm H}$ corresponding to the Lyman limit.
After a cell is ionized, its {\it residual} neutral fraction is determined assuming photo-ionization equilibrium:
\begin{equation}\label{eq:photoionizing_equilibrium}
	x_{\rm HI, res} f_{\rm ion,ss}\Gamma_{\rm ion} = \chi_{\rm HeII} \Delta\overline{n}_{\rm H} (1-x_{\rm HI, res})^2\alpha_{\rm B}
\end{equation}
where $\overline{n}_{\rm H}$ is the mean hydrogen number density while $\Delta$ is the cell's overdensity, $\chi_{\rm HeII}\sim1.08$
accounts for singly ionized helium,
$\alpha_{\rm B}$ is the case-B recombination coefficient, and $f_{\rm ion,ss}$ accounts for gas self-shielding {\color{black}(\citealt{Rahmati2013MNRAS.430.2427R}; see also \citealt{Chardin2018MNRAS.478.1065C})}. Note that sub-grid physics are implemented \citep{Sobacchi2014MNRAS.440.1662S} when calculating recombinations with the sub-grid density unresolved by our simulation cells assumed to follow a volume-weighted distribution of $P_{\rm V}(\Delta_{\rm sub}, z)$ from \citet{Miralda2000ApJ...530....1M}. However, we use the cell's mean overdensity when computing the Ly$\alpha$ optical depth{\color{black}, which neglects unresolved opacity fluctuations when calibrating to the hydrodynamic simulations below}. This will be improved in future work.

The IGM temperature ($T_{\rm g}$) is tracked following 
\citet{McQuinn2016MNRAS.456...47M}:
\begin{equation}\label{eq:tg}
T_{\rm g}^{\gamma} {=} T_{\rm ion,I}^{\gamma} \left[\left(\frac{\mathscr{Z}}{\mathscr{Z}_{\rm ion}}\right)^{3} \frac{n_{\rm H}}{n_{\rm H,ion}}\right]^{\frac{2\gamma}{3}} \frac{\exp\left(\mathscr{Z}^{2.5}\right)}{\exp\left(\mathscr{Z}_{\rm ion}^{2.5}\right)} {+} T_{\rm lim}^{\gamma}\frac{n_{\rm H}}{\overline{n}_{\rm H}},
\end{equation}
where we denote $\mathscr{Z}=(1+z)/7.1$ and use the subscript ``$_{\rm ion}$'' to indicate values at the time the cell was first ionized for convenience.
$\gamma = 1.7$ is the equation of state index while $T_{\rm lim}=1.8\mathscr{Z}\times10^{4}$~K \citep{Hui1997MNRAS.292...27H,Theuns1998MNRAS.301..478T,Puchwein2015MNRAS.450.4081P} and $T_{\rm ion,I}=2\times10^4$~K \citep{DAloisio2019ApJ...874..154D} are the relaxation and post I-front temperatures, respectively. Note that the scatter in $T_{\rm ion,I}$ has a negligible impact on the Ly$\alpha$ forest (e.g., \citealt{Davies2019MNRAS.489..977D}).

Finally, we compute the associated Ly$\alpha$ optical depth of each 1.95 cMpc simulation cell using a form of the Fluctuating Gunn-Peterson Approximation (FGPA; \citealt{gunn1965density,Weinberg1999elss.conf..346W}) for ionized cells:
\begin{equation}\label{eq:tau_alpha_GP}
	\tau_{\alpha, \rm GP} = \sqrt{\frac{3\pi\sigma_{\rm T}}{8}} f_{\alpha} \lambda_\alpha c H^{-1} n_{\rm H} x_{\rm HI}.
\end{equation}
Here $\sigma_{\rm T}$, $f_\alpha{=}0.416$ and $\lambda_\alpha{=}$1216{\AA} are the Thomson cross-section, Ly$\alpha$ oscillator strength, and Ly$\alpha$ rest-frame wavelength, respectively. Finally, we compute the effective optical depth, $\tau_{\rm eff, GP}$, following the same definition as the observation (see Section \ref{sec:obs}). 

The FGPA approximates the cross-section of Ly$\alpha$ absorption as a Dirac delta function at resonance and ignores peculiar velocities of the gas. In the following section we discuss how we use high-resolution hydrodynamic simulations to correct for the FGPA, accounting for missing small-scale structure.  This represents the main improvement of this work over our previous analysis in \citetalias{Qin2021MNRAS.506.2390Q}.

\subsection{Accounting for missing small-scale structure}
\label{subsec:calibration}

As mentioned above, our large-scale IGM simulations have a cell size of 1.95 cMpc.  This is a factor of few larger than the typical Jeans length in the ionized IGM and the width of the Ly$\alpha$ cross-section at resonance.  As a result, we use the FGPA in equation (\ref{eq:tau_alpha_GP}) instead of directly integrating over the full Voigt profile for the Ly$\alpha$ cross-section, $\sigma_{\alpha}$, and accounting for gas peculiar velocities:
\begin{equation}\label{eq:tau_alpha_full}
	\tau_{\alpha}= \int \frac{{\rm d}z}{1+z}  c H^{-1} n_{\rm H}x_{\rm HI} \sigma_{\alpha}.
\end{equation}
Does this approximation impact our modelled $\tau_{\rm eff}$ distributions?

The fact that $\tau_{\rm eff}$ is defined over $\Delta z =0.1$ (corresponding to roughly 20 of our IGM simulation cells) would suggest that this summary statistic is mostly sensitive to (resolved) large-scale fluctuations in flux.  
However, not resolving small-scale structures can effectively alias power towards large scales (e.g., \citealt{Viel2005PhRvD..71f3534V,Kooistra2022ApJ...938..123K}). 
Here we use a high-resolution hydrodynamic simulation from the Sherwood suite \citep{Bolton2017MNRAS.464..897B} to compare $\tau_{\rm eff, \rm GP}$ obtained from the low-resolution FGPA (equation \ref{eq:tau_alpha_GP}) against the correct calculation (equation \ref{eq:tau_alpha_full}).

We use a simulation with a cubic volume of $80h^{-1}$ cMpc on a side and $2\times512^3$ particles. It was performed using an updated version of \textit{Gadget}-2 \citep{Springel2005Natur.435..629S} and with a slightly different $\Lambda$CDM cosmology ($\Omega_{\mathrm{m}}, \Omega_{\mathrm{b}}, \Omega_{\mathrm{\Lambda}}, h, \sigma_8, n_s $ = 0.31, 0.048, 0.69, 0.68, 0.83, 0.96; \citealt{Planck2016A&A...594A..13P}). The modelled universe is exposed to a \citet{Haardt2012} UVB switched on at $z=15$. {\color{black}At $z\le5$, the simulated Ly$\alpha$ forest from Sherwood, which has a spatial resolution of <60 kpc,} agree very well with observational data \citep{Viel2013PhRvD..88d3502V,Bolton2017MNRAS.464..897B}.

\begin{figure*}[!ht]
	\centering
	\includegraphics[width=\textwidth]{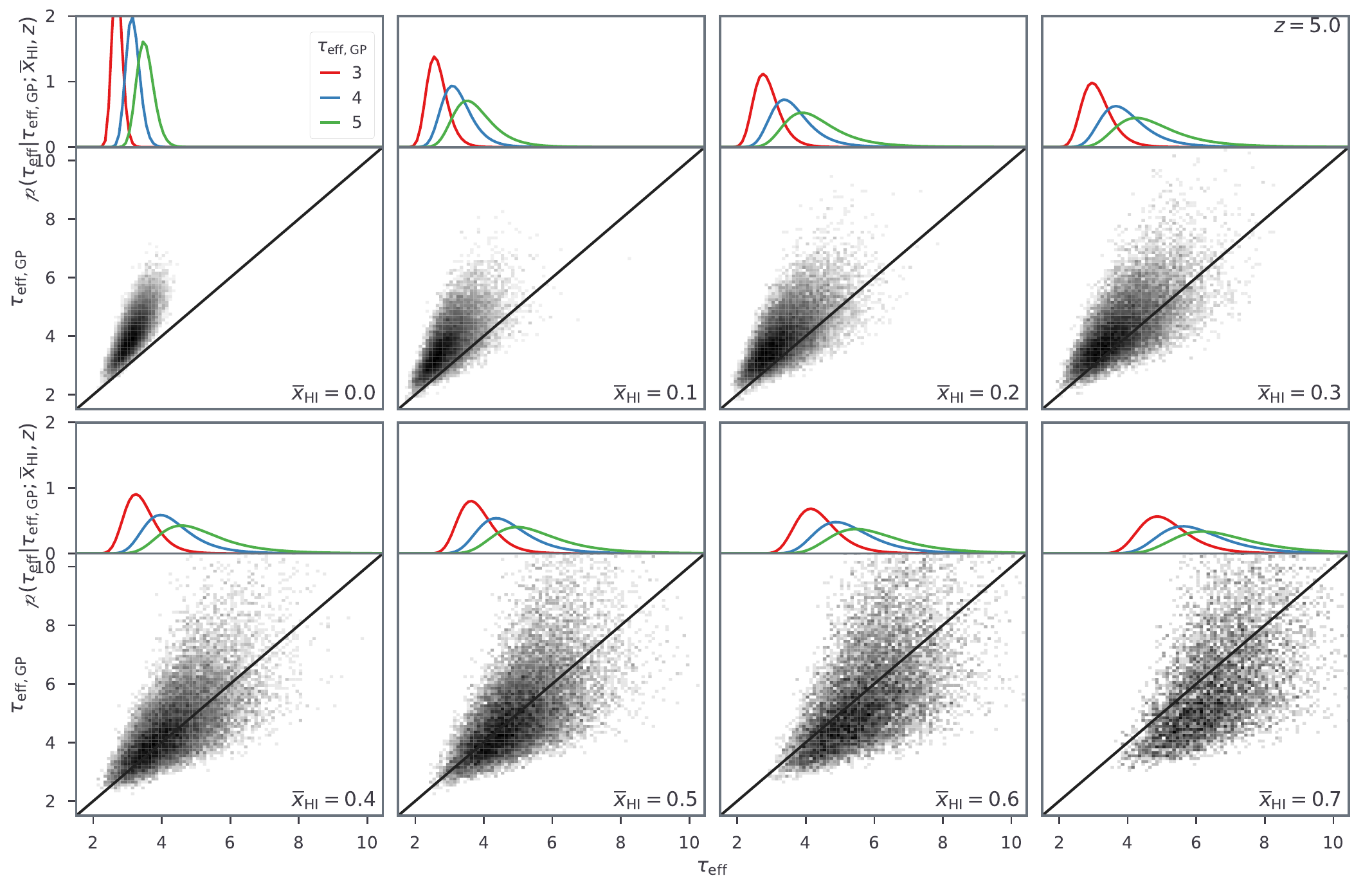}\vspace*{-3.4mm}
	\caption{{\color{black}{\it Lower sub-panels:} comparisons of the Ly$\alpha$ effective optical depth calculated using the full integral over the Ly$\alpha$ cross-section at the highest available resolution ($\tau_{\rm eff}$), to those calculated assuming the FGPA ($\tau_{\rm eff, GP}$). Both calculations use the Sherwood hydrodynamic simulation, with the latter obtained by down-sampling to the same low resolution adopted in our IGM forward-models and ignoring peculiar velocities. These sub-panels show pairs of $\tau_{\rm eff}$ -- $\tau_{\rm eff, GP}$ at different values of the mean neutral fraction, $\overline{x}_{\rm HI}$ --  an incomplete EoR is approximated by randomly placing spherical neutral patches in the simulation box until the desired filling factor of $\overline{x}_{\rm HI}$ is reached. These distributions of ($\tau_{\rm eff}$, $\tau_{\rm eff, GP}$) pairs are fit with KDE, resulting in a conditional probability distribution function $p(\tau_{\rm eff} ~|~ \tau_{\rm eff, GP}; \overline{x}_{\rm HI}, z)$, which is employed to correct our forward-modelled IGM lightcones for missing small scales. {\it Upper sub-panels}: example $\tau_{\rm eff}$ distributions conditioned at $\tau_{\rm eff, GP}=3$, 4 and 5.}
	\label{fig:tau_eff_compare}}
\end{figure*}

We project sightlines along each axis of the $z=5$ snapshot\footnote{Unfortunately, we did not have snapshots available at every redshift probed by observations, $z=5.1$, 5.2, 5.3...  We therefore perform our calibration only using the $z=5$ snapshot and assume the conditional distribution functions to be self-similar at other redshifts.  We crudely tested this assumption by scaling the density field by its mean evolution, finding only a $\tau_{\rm eff}\lesssim0.5$ shift in the resulting conditionals. {\color{black} We plan to improve the calibration in the future using more snapshots from higher resolution simulations.}}. This results in 5000 segments of length 80 $h^{-1}$ cMpc, which we bin to $\Delta z=0.1$. As the physical scale corresponding to $\Delta z=0.1$ changes with redshift, we repeat the binning for all redshifts spanned by the data, $z=5.1$, 5.2, 5.3, ..., 6.1.  For each bin, we compute the ``true'' effective optical depth ($\tau_{\rm eff}$; i.e. averaging the flux obtained using equation \ref{eq:tau_alpha_full}). We then recompute the effective optical depth of each segment assuming the same approximations we make in our large-scale IGM simulations (down-sampling the resolution and applying equation \ref{eq:tau_alpha_GP}) to obtain the corresponding $\tau_{\rm eff, GP}$.  We are then left with pairs of $\tau_{\rm eff}$ -- $\tau_{\rm eff, GP}$, which act as samples of the conditional probability of having a {\it true} $\tau_{\rm eff}$ given the corresponding FGPA value $\tau_{\rm eff, GP}\sim p(\tau_{\rm eff} ~|~ \tau_{\rm eff, GP}; z, \overline{x}_{\rm HI} = 0)$.

We then generalize this conditional probability to higher neutral fractions. Specifically,  we randomly place spherical neutral IGM patches in the Sherwood box until we obtain an HI filling factor of $\overline{x}_{\rm HI}$, repeating the above procedure to obtain $p(\tau_{\rm eff} ~|~ \tau_{\rm eff, GP}; z, \overline{x}_{\rm HI})$.  
We assume a log normal distribution peaked at a constant value of 4 cMpc for the radii of these HI patches. This is motivated by the results of \citet{Xu2017ApJ...844..117X}, who find a very modest evolution in the neutral patch size distribution during the final stages of the EoR. 

The resulting  $\tau_{\rm eff}$ -- $\tau_{\rm eff, GP}$ samples at $z=5$ are shown in Fig. \ref{fig:tau_eff_compare} where $\tau_{\rm eff}$ = $\tau_{\rm eff, GP}$ is marked by a diagonal line in each panel.  At low values of the neutral fraction, the FGPA tends to overestimate the true value of the effective optical depth.  This is especially evident at large overdensities with high values of $\tau_{\rm eff}$ \citep{Kooistra2022ApJ...938..123K}. As the neutral fraction increases, this bias decreases. However, the scatter in $p(\tau_{\rm eff} ~|~ \tau_{\rm eff, GP}; z, \overline{x}_{\rm HI})$ increases significantly. {\color{black}At $x_{\rm HI}\gtrsim0.5$ when damping-wing absorption becomes significant, the FGPA starts to instead underestimate the effective optical depth.}

We fit these samples with kernel density estimators (KDEs) in order to obtain an analytic form for $p(\tau_{\rm eff} ~|~ \tau_{\rm eff, GP}; z, \overline{x}_{\rm HI})$ that can be evaluated when forward modelling (c.f. Fig. \ref{fig:reference}).
Specifically, we use 2D conditional Gaussian distributions from the {\tt conditional\_kde}\footnote{\url{https://github.com/dprelogo/conditional_kde}} package to fit the samples of $1/\tau_{\rm eff}$ -- $1/\tau_{\rm eff, GP}$ (as the reciprocal of the optical depth more closely follows a Gaussian distribution).  The parameters of the {\color{black}Gaussian kernels} (means {\color{black}and standard deviations}) are explicit functions of redshift and the neutral fraction\footnote{Throughout this paper, we list the fitted functional dependencies of probability distributions to the right of a semi-colon.  Thus $p(\tau_{\rm eff} ~|~ \tau_{\rm eff, GP}; z, \overline{x}_{\rm HI})$ is a conditional probability of $\tau_{\rm eff}$ given $\tau_{\rm eff, GP}$, whose parameters (mean and sigma) are functions of $z$ and $\overline{x}_{\rm HI}$.}, allowing us to easily evaluate $p(\tau_{\rm eff} ~|~ \tau_{\rm eff, GP}; z, \overline{x}_{\rm HI})$ at any neutral fraction and redshift.
We show some examples of the fitted conditional distributions in the upper sub-panels of Fig. \ref{fig:tau_eff_compare}.

\subsection{Computing the forest likelihood}\label{subsec:likelihood_forest}

For each $\tau_{\rm eff, GP}(\overline{x}_{\rm HI}, z)$ calculated using equation (\ref{eq:tau_alpha_GP}) on our IGM lightcones, we obtain a {\color{black} random} sample from the conditional distributions discussed in the previous subsection: $\tau_{\rm eff} \sim p(\tau_{\rm eff} ~|~ \tau_{\rm eff, GP}; z, \overline{x}_{\rm HI})$.  
Therefore, this leads to a set of effective optical depths that are stochastically corrected for missing sub-structure {\color{black}in the FGPA method}. We additionally account for uncertainty in the continuum reconstruction by adding $\ln(\mathcal{R})$ to every $\tau_{\rm eff}$ sample.
 Here, $\mathcal{R}$ is a random number following a normal distribution centred at unity with a standard deviation of 10\%, typical of the continuum reconstruction relative errors \citep{Bosman2022MNRAS.514...55B}.  
 Note that we do not account for wavelength correlations in the reconstruction errors {\color{black} or the actual ``usable'' range of observed wavelengths in each quasar spectrum (see more in \citealt{Bosman2022MNRAS.514...55B})}; we plan on including these in future work.  We fit the resulting histograms of $\tau_{\rm eff}$ in each of the redshift bins defined by the data to obtain the PDFs, $p(\tau_{\rm eff}; z)$.

These PDFs are our theoretical expectation of the real Universe, for a given model and choice of astrophysical parameters. Therefore, each {\it observed} value of $\tau_{\rm eff}^i$ at $z^i$  corresponds to a sample from the theoretical PDF, with a corresponding likelihood $p(\tau_{\rm eff} = \tau_{\rm eff}^i ; z=z^i)$.  
For non-detections, we take $\tau_{\rm eff}^i$ to be the 2 $\sigma$ lower limit implied by the noise \citep{Bosman2022MNRAS.514...55B}.

It is worth noting that the likelihood distribution, $p(\tau_{\rm eff},z)$, is forward-modelled {\color{black}(i.e., it is sampled by running a simulator many times)}, without having to explicitly adopt a functional form.  This is referred to as an {\it implicit} likelihood.
Inferences using an implicit likelihood (also called simulation-based inference) are becoming increasingly popular in this field (e.g., \citealt{Zhao2022ApJ...933..236Z,Prelogovic2023MNRAS.524.4239P,Davies2024ApJ...965..134D,Greig2024MNRAS.533.2502G,Greig2024MNRAS.533.2530G}) as most EoR datasets do not have an analytically-tractable likelihood; and common assumptions of Gaussian pseudo-likelihoods can result in biased posteriors  (see \citealt{Prelogovic2023MNRAS.524.4239P}).

We obtain the final forest likelihood by multiplying the implicit likelihoods over all XQR-30+ quasars, $i$, and over all redshift bins used in the analysis.  Specifically, we take $\mathcal{L}_{\rm forest}=\prod_{z=5.3}^{z=6.1} \prod_{i} p(\tau_{\rm eff} = \tau_{\rm eff}^i ; z)$.  We do not include data at $z\leq5.2$ in order to make our likelihood more sensitive to the EoR (see e.g., \citealt{Bosman2022MNRAS.514...55B} who showed that the EoR ends sometime before $z\sim5.2$).

Note that this procedure does not account for higher order correlations in the mapping from $\tau_{\rm eff, GP}$ to $\tau_{\rm eff}$.  Moreover, it assumes that each $\Delta z$ = 0.1 ($\sim$ 40 cMpc) segment is an independent sample of $p(\tau_{\rm eff}; z)$; i.e. we ignore the covariance between the $\Delta z$ = 0.1 segments extracted from a single quasar spectrum.  We expect the covariance on such large scales to have only a minor impact on the total likelihood.  Nevertheless, we plan on relaxing this approximation in future work in which we will use simulation-based inference for the total likelihood, accounting for large scale correlations in both the effective optical depths and reconstruction errors.

\subsection{Combining with complementary  observations}\label{subsec:likelihood_other}

We also account for complementary, independent data when performing inference.  Specifically, we compute additional likelihood terms for: (i) the galaxy non-ionizing UV LFs well-established by {\it Hubble} at $6\leq z \leq 10$ \citep{Bouwens2015ApJ...803...34B,Bouwens2016ApJ...830...67B,Oesch2018ApJ...855..105O}; and (ii) the CMB polarization power spectra observed by {\it Planck} \citep{Planck2020A&A...641A...6P}.  These two datasets are independent and mature, and can therefore be interpreted robustly.  Unlike  \citetalias{Qin2021MNRAS.506.2390Q}, we do not include a likelihood term for the pixel Dark Fraction \citep{Mesinger2010MNRAS.407.1328M,McGreer2015MNRAS.447..499M} as this statistic is also based on Lyman forests and therefore is technically not fully independent from the optical depth distributions discussed above. Thus our total likelihood consists of the product of three terms: $\mathcal{L}_{\rm tot}~=~\mathcal{L}_{\rm forest } \times \mathcal{L}_{\rm LF} \times \mathcal{L}_{\rm CMB}$, where the final two correspond to the LF and CMB likelihoods, discussed further below.

We construct the UV LF likelihood following \citet{Park2019MNRAS.484..933P}.  Specifically, we assume a Gaussian likelihood in each magnitude bin, $M_{\rm UV,i}$, with a negligible covariance between bins (see e.g., \citealt{Leethochawalit2022arXiv220515388L} for an alternative approach).  The UV LF likelihood is thus $\mathcal{L}_{\rm LF,tot}=\prod_{z=6}^{z=10} \prod_{i} \exp\left\{-\left[{\Delta \phi(M_{\rm UV,i})}/{\sigma_\phi(M_{\rm UV,i})}\right]^2\right\}$, where $\Delta \phi$ is the  difference between forward-modelled and observed  galaxy number densities in a given magnitude and redshift bin, and the corresponding observational uncertainties are $\sigma_\phi$.  We only consider magnitudes fainter than $M_{\rm UV}=-20$ to avoid modelling dust attenuation, and use the redshift range between $z=6$ and 10 spanning the EoR.

We construct the {\it Planck} CMB likelihood as a two-sided Gaussian on the Thomson scattering optical depth summary statistic inferred by \citet{Qin2020MNRAS.499..550Q}: $\tau_e=0.0569^{+0.0073}_{-0.0066}$.  Specifically, we take the form $\mathcal{L}_{\rm CMB} = \exp\left[-\left({\Delta \tau_e}/{\sigma_{\tau_e}}\right)^2\right]$, where $\Delta \tau_e$ represents the difference between the forward-modelled and measured optical depths while $\sigma_{\tau_e}$ is the observational uncertainty.  Note that \citet{Qin2020MNRAS.499..550Q} found very little difference in the inferred posteriors when using a likelihood defined directly on the E-mode polarization power spectra compared to using a Gaussian likelihood on the $\tau_e$ summary derived from the power spectra.  We thus use the latter as it is much more computationally efficient.

\subsection{Summary of model parameters and associated priors}\label{sec:inference}

Before showing our inference results, we summarize the free parameters used in our galaxy models and their associated prior ranges (see also Table \ref{tab:source_model}).

\begin{enumerate}[labelwidth=!,itemindent=24pt,labelindent=0pt, leftmargin=0em, itemsep=3pt, parsep=0pt, topsep = 3pt]
\item $\log_{10}f_{*,10}\in[-2,-0.5]$: the fraction of galactic baryons in stars, normalized at $M_{\rm vir}=10^{10} M_\odot$.  This parameter sets the normalization of the stellar-to-halo mass relation, and its prior range is motivated by observations and simulations of high-redshift galaxies \citep{Dayal2014MNRAS.445.2545D,Mutch2016,Behroozi2019MNRAS.488.3143B,Bird2022MNRAS.512.3703B,Stefanon2021ApJ...922...29S}.
\item $\alpha_*\in[0, 1.0]$: the power law index relating the stellar fraction to the halo mass.  This parameter determines the slope of the stellar-to-halo mass relation.  Observations of the faint end of the UV LFs suggest more efficient star formation in more massive galaxies \citep{Bouwens2015ApJ...803...34B,Oesch2018ApJ...855..105O}, motivating our prior range.
\item $\log_{10}f_{\rm esc,10}\in[-3,0]$: the amplitude of the power-law relating the UV ionizing escape fraction to halo mass, normalized at $M_{\rm vir}=10^{10} M_\odot$.  The wide prior reflects the large uncertainties in both low-redshift observations (\citealt{Vanzella2010ApJ...725.1011V,Vanzella2016ApJ...825...41V,Boutsia2011ApJ...736...41B,Nestor2013ApJ...765...47N,Guaita2016A&A...587A.133G,Grazian2016A&A...585A..48G,Shapley2016ApJ...826L..24S,Bian2017ApJ...837L..12B,Steidel2018ApJ...869..123S,Naidu2018MNRAS.478..791N,Fletcher2019ApJ...878...87F,Izotov2021MNRAS.503.1734I,Pahl2021MNRAS.505.2447P}) and reionization simulations (\citealt{Kostyuk2023MNRAS.521.3077K,Choustikov2024MNRAS.529.3751C,Mutch2024MNRAS.527.7924M}). 
\item $\alpha_{\rm esc}\in[-1, 0.5]$: the power law slope of the UV ionizing escape fraction to halo mass relation. Galaxy simulations seem to suggest boosted Lyman continuum leakage in less massive galaxies {\color{black}as supernovae evacuate low column density channels from shallow gravitational potentials} \citep{Paardekooper2015MNRAS.451.2544P,Xu2016ApJ...833...84X,Kostyuk2023MNRAS.521.3077K,Mutch2024MNRAS.527.7924M}. This motivates a wider negative range in the prior, although we caution that this is highly uncertain and therefore still allow positive values in our prior {\color{black}(e.g., \citealt{Ma2015MNRAS.453..960M,Naidu2020ApJ...892..109N,Rosdahl2022MNRAS.515.2386R,Bhagwat2024MNRAS.531.3406B})}.
  \item $\beta_{\rm esc}\in[-3, 3]$: the power law scaling index of the UV ionizing escape fraction as a function of redshift, used only in $\atomicseven$. The prior is somewhat arbitrary with the upper and lower limits allowing $f_{\rm esc}$ to scale similarly\footnote{Because of the wide prior, we save computational time by initially performing a fast, approximate likelihood estimate, which shows that the posterior peaks at negative values of $\beta_{\rm esc}$.  We then perform our fiducial inference on a narrower prior range  $\beta_{\rm esc}{\in}[-3.0, 0]$ to save computational overheads, but scale the Bayesian evidence to account for the missing prior volume (see e.g., \citealt{Murray2022MNRAS.517.2264M}).}.
\item $\tau_\ast\in(0, 1]$: the star formation timescale in units of the Hubble time.
The flat prior encompasses extreme cases where the entire stellar mass is formed in an instantaneous burst event or gradual built over the age of the universe. 
\item $\log_{10}\left(M_{\rm turn}/\Msun\right)\in[8, 10]$: the characteristic halo mass  below which star formation becomes exponentially suppressed. The lower and upper limits of the flat prior are motivated by the atomic cooling threshold and the faintest, currently observed high-redshift galaxies (e.g., \citealt{Bouwens2015ApJ...803...34B,Bouwens2016ApJ...830...67B,Oesch2018ApJ...855..105O}), respectively.
\end{enumerate}

\section{Fiducial inference results}\label{sec:result}

As can be seen from Table  \ref{tab:source_model}, the Bayesian evidence ratios suggest that the data have a {\it very strong} preference (\citealt{Jeffreys1939thpr.book.....J}) for the $\atomicseven$ model.  We therefore treat this model as ``fiducial'', presenting its posterior in this section, before comparing it to the  $\atomicsix$ model in the following section.\footnote{Note that having a much higher Bayesian evidence does not necessarily mean that the model is ``the correct'' one.  Compared to the alternate model, the fiducial one is more flexible and predictive given the observed data, without wasted prior volume.}
Alternatively, one could do Bayesian model averaging to combine the derived IGM and galaxy properties from different models; however the evidence ratio in this case is so strongly skewed towards $\atomicseven$, that the model-averaged posteriors would just follow the  $\atomicseven$ ones.

We show the posteriors in the space of galaxy parameters in \ref{app:paramter_posterior}.  Here we focus on the inferred IGM properties and galaxy scaling relations.

\subsection{Effective optical depth distributions}

\begin{figure*}[!ht]
	\begin{minipage}{\textwidth}
		\centering
		\includegraphics[width=0.92\textwidth]{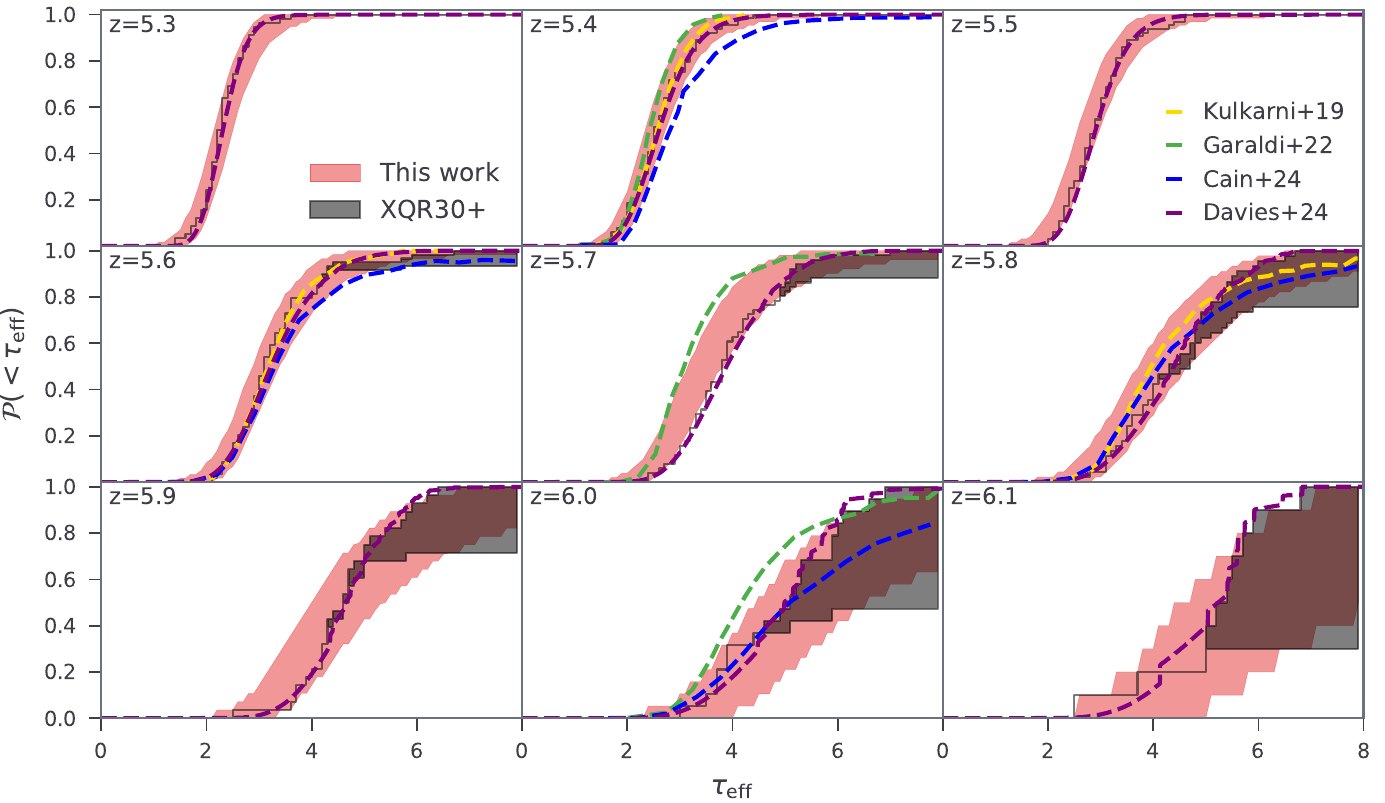}\vspace*{-1.3mm}
		\caption{Inferred $\tau_{\rm eff}$ CDFs from our fiducial model ({\it red}) from $z=$5.3 to 6.1.  To account for cosmic variance, we randomly select from each model in the posterior the same number of sightlines as in the XQR-30+ observational dataset.  The red regions indicate the 95\% C.I. For comparison, the XQR-30+ observations are shown in grey with non-detections denoted with the shaded regions spanning the flux range between zero and double the noise \citep{Bosman2022MNRAS.514...55B}. A number of theoretical results are shown for comparison (\citealt{Kulkarni2019MNRAS.485L..24K,Garaldi2022MNRAS.512.4909G,Cain2024MNRAS.531.1951C}{\color{black}; \citealt[][optimistic]{Davies2024ApJ...965..134D};} with earlier works using {\color{black}slightly different} binning for $\tau_{\rm eff}$). 
		}\label{fig:post_CDF}
	\end{minipage}
\end{figure*}

In Fig. \ref{fig:post_CDF} we plot the recovered optical depth CDFs in red enclosing the 95\% confidence interval (C.I.). Observational data from \citet{Bosman2022MNRAS.514...55B} are shown in grey.  The red shaded regions are constructed from the posterior samples, each having the same number of randomly-selected sightlines per redshift bin as in the data to account for cosmic variance.   We note that the cosmic variance dominates the widths of the CDFs, especially at the highest redshifts.

We see from the figure that our fiducial model excels at recovering the observed $\tau_{\rm eff}$ CDFs throughout this redshift range -- despite individual $\tau_{\rm eff}$ data being used in the likelihood, our model can recover both the mean {\it and} the shape of the observed optical depth distribution. We stress that most previous work either used hyperparameters to account for the mean opacity evolution (e.g., \citetalias{Qin2021MNRAS.506.2390Q}), calibrated the models to have the same mean opacity as the data and/or treated each redshift bin independently (e.g., \citealt{Kulkarni2019MNRAS.485L..24K,Meiksin2020MNRAS.491.4884M,Cain2024MNRAS.531.1951C}, \citealt{Gaikwad2023MNRAS.525.4093G}\footnote{{\color{black}In order to constrain the MFP and UVB using Kolmogorov–Smirnov test statistics, \citet{Gaikwad2023MNRAS.525.4093G} treated non-detections in slightly different ways when calculating the CDFs from data and their model.}}, \citealt{Davies2024ApJ...965..134D}).  Some more expensive coupled hydrodynamic and radiative-transfer simulations such as CODA and THESAN (\citealt{Ocvirk2021MNRAS.507.6108O,Garaldi2022MNRAS.512.4909G}) do not directly tune their simulations to reproduce the forest data; however their predicted CDFs do not agree with the data as well as most of the other previously-mentioned works.
For illustration, we show some of these results in Fig. \ref{fig:post_CDF}.

\subsection{EoR history}
\label{sec:eor}

\begin{figure*}[!ht]
	\begin{minipage}{\textwidth}
		\centering
		\includegraphics[width=.78\textwidth]{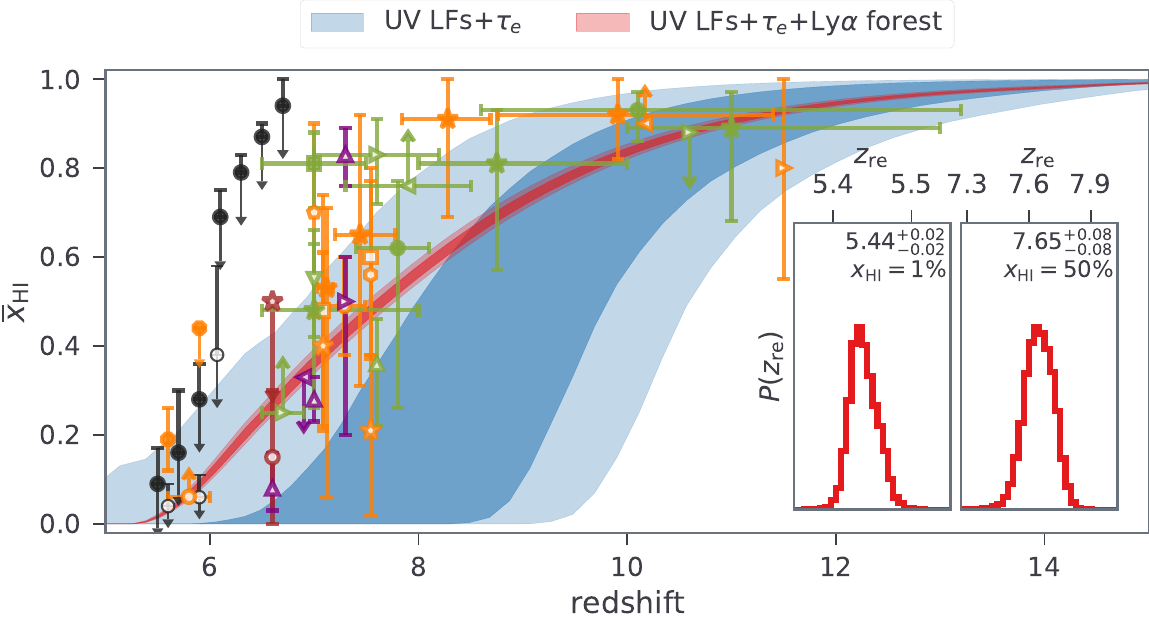} \vspace*{-2.5mm}
		\caption{The inferred EoR history using our fiducial model.  The blue shaded region uses only UV LFs and CMB $\tau_e$ data (a likelihood of $\mathcal{L}_{\rm LF} \times \mathcal{L}_{\rm CMB}$), while the red  additionally includes the Ly$\alpha$ forest $\tau_{\rm eff}$ distributions (likelihood of $\mathcal{L}_{\rm forest } \times \mathcal{L}_{\rm LF} \times \mathcal{L}_{\rm CMB}$).  In both cases the dark (light) regions indicate the 68\% and 95\% C.I.  The XQR-30+ forest data are very constraining; including them makes the posterior transition from being prior-dominated to being likelihood-dominated. PDFs of the redshifts corresponding to $\overline{x}_{\rm HI}=$ 0.01 and 0.5 are presented in the inset panels, showing that in our fiducial model reionization ends at $z=5.44\pm0.02$ and the EoR mid-point is at $z=7.7\pm0.1$. Estimates of the ionization state of the universe coming from other probes are also shown for illustrative purposes including the dark pixel upper limits (black; \citealt{McGreer2015MNRAS.447..499M,Jin2023ApJ...942...59J}), Lyman-$\alpha$ damping-wing absorption in QSOs (orange; \citealt{Mesinger2004ApJ...611L..69M,Banados2018Natur.553..473B,Greig2017,Greig2019,Greig2022,Davies2018,Wang2020ApJ...896...23W,Spina2024A&A...688L..26S,Zhu2024MNRAS.533L..49Z}) or in galaxies (orange; \citealt{Curtis-Lake2023NatAs...7..622C,Hsiao2024ApJ...973....8H,Umeda2024ApJ...971..124U}), Lyman-$\alpha$ equivalent widths  (green; \citealt{Mesinger2015MNRAS.446..566M,Mason2019MNRAS.485.3947M,Jung2020ApJ...904..144J,Whitler2020MNRAS.495.3602W,Bolan2022MNRAS.517.3263B,Bruton2023ApJ...949L..40B,jones2024jadesmeasuringreionizationproperties,Nakane2024ApJ...967...28N,Tang2024ApJ...975..208T}), and the LF or clustering of Ly$\alpha$ emitters (purple; \citealt{Sobacchi2015MNRAS.453.1843S,Inoue2018PASJ...70...55I,Ouchi2018PASJ...70S..13O,Morales2021ApJ...919..120M,Wold2022ApJ...927...36W,}), most of which are consistent with our results despite not being included in the inference. 
  \label{fig:post_xH}
  		}
    
	\end{minipage}
\end{figure*}

In Fig. \ref{fig:post_xH} we show the main result of this work -- the inferred reionization history in our fiducial model.   In blue we show the posterior resulting from using only the $\mathcal{L}_{\rm LF}$ and $\mathcal{L}_{\rm CMB}$ likelihood terms.  This roughly corresponds to our previous state of knowledge, without using the forest data.\footnote{The literature has many additional estimates of the EoR history that we do not include in our inference (see for instance data points shown in Fig. \ref{fig:post_xH}). 
As mentioned above, interpreting these observations is very challenging and prone to observational and modelling systematics.  Robust interpretation would require dedicated forward-models of each observation and associated systematics.  In any case, these alternate probes only weakly constrain the EoR history using current data (\citealt{Mesinger2004ApJ...611L..69M,Greig2022,Bruton2023ApJ...949L..40B,Ouchi2018PASJ...70S..13O,Reichardt2021ApJ...908..199R}).  We therefore expect our results to not be impacted by the inclusion of additional datasets.} 
From the blue region we see that the CMB optical depth and the UV LFs do not result in tight constraints on the EoR history.   The UV LFs loosely constrain the evolution of the star formation rate density (SFRD), while the CMB optical depth additionally constrains the corresponding ionizing escape fraction (see the parameter posterior shown in blue in Fig. \ref{fig:post_paramter}).  Given that these constraints are not tight, the posterior is prior dominated (as opposed to being likelihood dominated).  Since we chose broad priors, allowing the ionizing escape fraction to extend to unity, most of the posterior volume traces a relatively early reionization, with midpoints around $z=8$--9.

The red shaded region in the figure shows what happens to the posterior when we further include the Ly$\alpha$ forest data, i.e. with the total likelihood of $\mathcal{L}_{\rm forest } \times \mathcal{L}_{\rm LF} \times \mathcal{L}_{\rm CMB}$.  The EoR history, $\overline{x}_{\rm HI}(z)$, of the maximum-a-posteriori (MAP) model is listed in Table \ref{tab:xhi_gamma_mfp}, and is well fit by a rational function
\begin{equation}\label{eq:rational}
\begin{split}   
    f(z) = \frac{m_0+m_1z+m_2z^2+m_3z^3}{n_0+n_1z+n_2z^2+n_3z^3}~,
\end{split}
\end{equation}
\noindent with parameters $\{m_0,m_1,m_2,m_3,n_0,n_1,n_2,n_3\}$ = \{292.6, -105.47, 7.824, 0.312, -24.3, 22.9, -4.96, 0.694\}. It is obvious that the $\tau_{\rm eff}$ data are extremely constraining, resulting in a very narrow posterior.  The uncertainties are over an order of magnitude smaller than without the forest data, with most of the history constrained to better than $\Delta z \sim 0.1$ at the 68\% C.I.
The forest data {\it require} the EoR to be ongoing below $z \leq 6$ (see also the previous results in \citealt{Choudhury2020arXiv200308958C} and \citetalias{Qin2021MNRAS.506.2390Q}).  From the inset panels in the figure, we see that in this fiducial model reionization ends at $z=5.44\pm0.02$ and the EoR mid-point is at $z=7.7\pm0.1$. {\color{black} Consequently, the inferred CMB optical depth is also tightly constrained with $\tau_{e}=0.0589\pm 0.001$ ($1\sigma$) compared to $\tau_{e}=0.0571\pm 0.006$ when the forest is not included.}

Perhaps surprisingly, the forest data tightly constrain the EoR history at redshifts beyond where we have forest data, $z>6.3$.  These constraints are {\it indirect}, coming from the combination of HMF evolution, the SFR to halo mass implied by UV LF observations, and the ionizing escape fraction scalings required to match the forest.  The forest in particular provides a firm anchor for our models. {\color{black}The forest data} requires a photon-starved end to reionization, with recombinations starting to balance ionizations, in order to smoothly transition into the post EoR regime \citep{Bolton2007MNRAS.382..325B, Sobacchi2014MNRAS.440.1662S}.  Such a ``soft-landing'' is difficult to achieve with small-box EoR simulations (e.g., \citealt{Barkana2004ApJ...609..474B}) and/or with those that cannot resolve recombinations in the late EoR stages (c.f. Fig. 6 in \citealt{Sobacchi2014MNRAS.440.1662S}, and \citealt{Qin2021MNRAS.506.2390Q, Cain2024MNRAS.531.1951C}).  Such  limitations tend to result in an overly rapid evolution of the late EoR stages, which in turn requires ad-hoc corrections/tuning (e.g., a very rapid drop in emissivity) in order to match forest data {\color{black}(see Fig. \ref{fig:rec} and associated discussion)}.  The fact that our box sizes are 250 Mpc and that sub-grid recombinations are computed analytically (and thus not limited by resolution), likely allows us to capture this ``soft-landing'' preferred by the forest data.  We caution however that these constraints on the EoR history at $z>6.3$ are indirect, and as such become increasingly model-dependent at increasingly higher redshifts.  We will revisit this in the future using alternate galaxy models that include an additional population of early, molecular-cooling galaxies, which might dominate the ionizing background at $z>10$--15 (e.g., \citealt{Qin2020MNRAS.495..123Q,Munoz2022MNRAS.511.3657M,Ventura2024MNRAS.529..628V}).

For illustration, we also plot various independent estimates of the IGM neutral fraction, using other probes.  These typically come from the analysis of IGM damping wing absorption in QSO or galaxy spectra  (see the caption of Fig. \ref{fig:post_xH} for details).
Our EoR posterior is qualitatively consistent with most of these estimates, despite not including them in our likelihood. This lends confidence that damping wing analysis can be reasonably trusted, despite the associated systematics and modelling challenges (e.g., \citealt{Mesinger2004ApJ...611L..69M,Banados2018Natur.553..473B,Greig2022,Wang2020ApJ...896...23W,Zhu2024MNRAS.533L..49Z,Hennawi2024arXiv240612070H,Kist2024arXiv240612071K}).

\subsection{UVB and MFP evolution}

\begin{figure}[!ht]
	\begin{minipage}{\columnwidth}
		\centering
		\includegraphics[width=\textwidth]{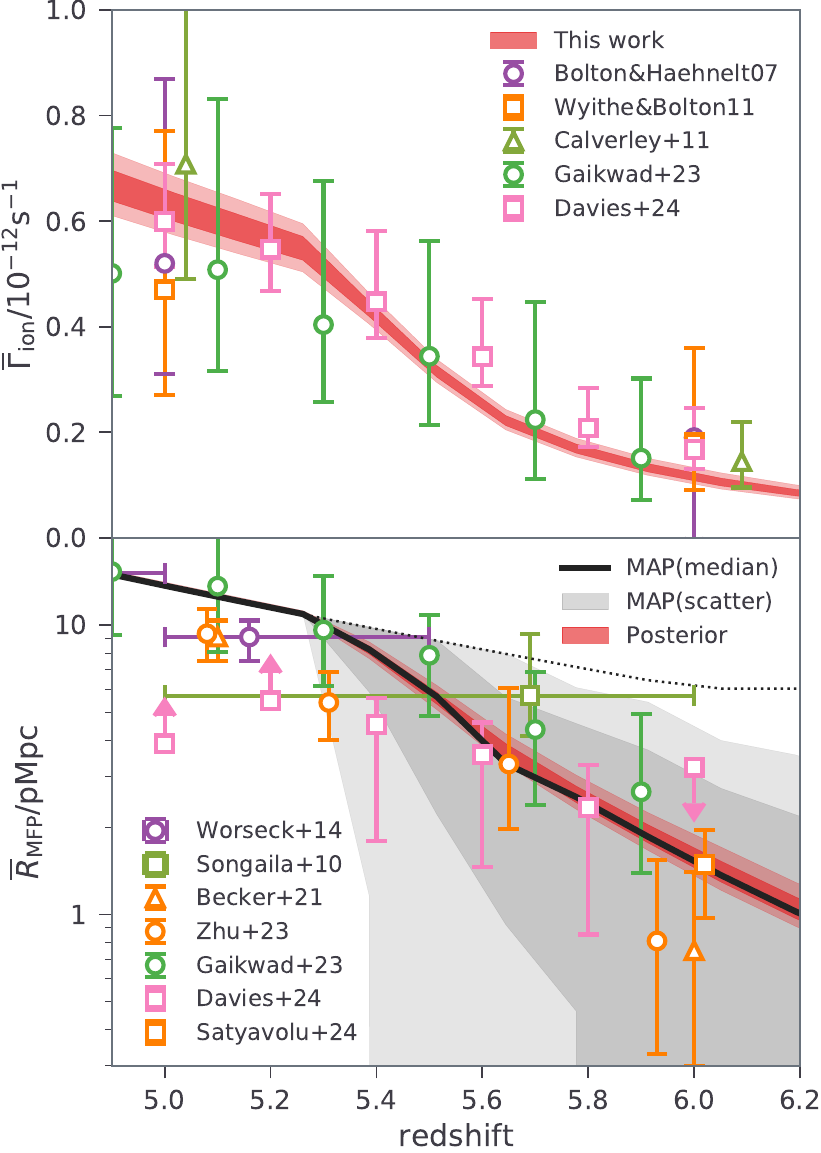}
		\caption{The posterior of our fiducial model in the space of the mean photo-ionization rate ({\it top panel}) and proper mean free path ({\it bottom panel}). 
  As in the previous figure, the dark (light) shaded region corresponds to 68\% (95\%) C.I. {\color{black}In the lower panel, we additionally show the volume distribution of the MFPs from the MAP model (median and scatters). The dotted line indicates the assumed $R_{\rm MFP, LLS}$.}
  Various previous estimates from the forests \citep{Bolton2007MNRAS.382..325B,Wyithe2011MNRAS.412.1926W,Calverley2011MNRAS.412.2543C,Worseck2014MNRAS.445.1745W,Songaila2010ApJ...721.1448S,Becker2021MNRAS.508.1853B,Gaikwad2023MNRAS.525.4093G,Zhu2023ApJ...955..115Z,Davies2024ApJ...965..134D,Satyavolu2024MNRAS.533..676S} are also shown with their 68\% error bars.
  Our results are in general agreement with these independent estimates, despite not having used them in the inference.
  \label{fig:post_global}
  		}
    
	\end{minipage}
\end{figure}

Fig. \ref{fig:post_global} shows the inferred redshift evolution of the photo-ionization rate and MFP in our fiducial model.  The forest data are able to constrain these global IGM quantities at percent level precision.  The total MFP converges to our assumed uniform value for the ionized IGM post EoR at $z\lesssim 5.2$ (i.e. $R_{\rm MFP,LLS}$).  Neutral patches during the EoR contribute increasingly to the MFP at earlier stages, as discussed in Sec. \ref{subsec:igm_sim} (see also \citealt{Roth2024MNRAS.530.5209R}).  This results in a more rapid drop in the MFP  from $z\sim5$ to 6 than would be expected in simple, uniform-UVB, post-EoR models (e.g., \citealt{Becker2021MNRAS.508.1853B}).

In the figure we also show several independent estimates from the literature. These come from: (i) adjusting simulated Ly$\alpha$ optical depths to match the observed flux evolution (e.g., \citealt{Bolton2007MNRAS.382..325B}; (ii) estimating the column-density evolution of HI absorbers \citep{Songaila2010ApJ...721.1448S}; (iii) modelling the size evolution of quasar near zones \citep{Wyithe2011MNRAS.412.1926W}; (iv) modelling flux profiles around near zones \citep{Calverley2011MNRAS.412.2543C, Worseck2014MNRAS.445.1745W, Becker2021MNRAS.508.1853B,Zhu2023ApJ...955..115Z,Satyavolu2024MNRAS.533..676S}; and (v) co-varying the MFP and UVB to match forest fluctuations independently at each redshift \citep{Davies2024ApJ...965..134D,Gaikwad2023MNRAS.525.4093G}.
Our results are generally in good agreement with these independent estimates, despite the fact that we do not use them in our analysis. Our recovered MFP at $z\sim6$ is on the upper end of the 68\% error bars from \citet{Becker2021MNRAS.508.1853B,Zhu2023ApJ...955..115Z}. 
This mild tension could point to additional systematics in these observational interpretations {\color{black}(see more in \citealt{Satyavolu2024MNRAS.533..676S})} and/or missing physics in our models, such as gas relaxation (e.g., \citealt{Park2016ApJ...831...86P, DAloisio2020ApJ...898..149D}) or the inhomogeneous post I-front temperature (e.g., \citealt{DAloisio2019ApJ...874..154D, Davies2019MNRAS.489..977D}).  We plan on investigating these effects in future work.

{\color{black}In the bottom panel of Fig. \ref{fig:post_global}, we additionally present the volume distribution of the MFPs derived from our MAP model. The black curve represents the median while the grey shaded regions indicate 68\% and 95\% of the volume distribution.  We see that 68\% of the volume has an MFP determined by $R_{\rm MFP, EoR}$ at $z\sim5.5$, even though reionization completes at $z=5.44$. This finding underscores that the assumed functional form of $R_{\rm MFP, LLS}$ likely has a minor impact on the MFP at these EoR redshifts. Nevertheless in future work we will additionally sample the uncertainties in the mean and scatter of $R_{\rm MFP, LLS}$.}

\subsection{UV ionizing emissivity}

\begin{figure}[!ht]
	\begin{minipage}{\columnwidth}
		\centering
		\includegraphics[width=\textwidth]{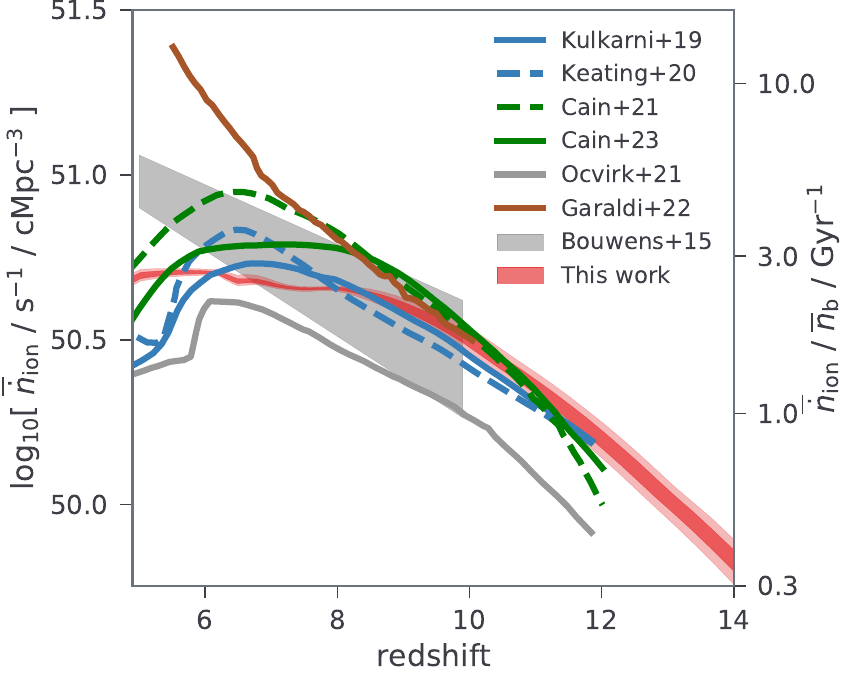}\vspace*{-1.3mm}
		\caption{The inferred UV ionizing emissivity. {\color{black} On the left axis we denote the number of ionizing photons per time per comoving volume, while on the right axis we show the number of ionizing photons per time per baryon.} As in the previous figure, the dark (light) shaded red region corresponds to 68\% (95\%) C.I.  For comparison, we include other estimates from: (i) simulations tuned to match the forest opacity distributions \citep{Kulkarni2019MNRAS.485L..24K,Keating2020MNRAS.491.1736K,Cain2021ApJ...917L..37C}; (ii) coupled hydrodynamic and radiative-transfer simulations \citep{Garaldi2022MNRAS.512.4909G, Ocvirk2021MNRAS.507.6108O}; and (iii) a simple empirical relation based on assuming a constant escape fraction and SFRD extrapolated down to a fixed limiting magnitude of $M_{\rm UV}=-13$ \citep{Bouwens2015b}.
  \label{fig:post_emissivity}
  		}
	\end{minipage}
\end{figure}

In Fig. \ref{fig:post_emissivity}, we present the inferred ionizing emissivity evolution in our fiducial model. Unlike many previous studies (e.g., \citealt{Kulkarni2019MNRAS.485L..24K,Keating2020MNRAS.491.1736K,Cain2021ApJ...917L..37C}), we reproduce the Ly$\alpha$ opacity distribution without requiring a sharp drop in the emissivity at $z\lesssim7$.  Such a rapid drop in the emissivity would be difficult to reconcile with the more gradual evolution implied by observations of galaxy UV LFs (e.g., \citealt{Bouwens2015b}), as it requires either fast evolving feedback in faint galaxies (e.g., \citealt{Ocvirk2021MNRAS.507.6108O}; though see e.g., \citealt{Sobacchi2014MNRAS.440.1662S, Katz2019arXiv190511414K}) or in their ionizing escape fractions.  Even under both such putative scenarios, it is difficult to physically justify cosmological evolution that is more rapid than characteristic time-scales during this epoch which are generally $\gtrsim$ 200 Myrs (e.g., the duration of the EoR, halo dynamical and/or sound crossing times; c.f. \citealt{Sobacchi2013MNRAS.432.3340S}).  As discussed in Section \ref{sec:eor}, one possible explanation is that simulating the end stages of the EoR requires very large-scale and very high-resolution hydrodynamic simulations to track the rapid evolution of self-shielding in the IGM and the strong spatial correlation between ionizing sinks and sources.  Our calibrated sub-grid approach could allow us to capture the relevant recombination physics without requiring very high resolution (\citealt{Sobacchi2014MNRAS.440.1662S}).

\begin{figure*}[!ht]
	\begin{minipage}{\textwidth}
		\centering
		\includegraphics[width=.9\textwidth]{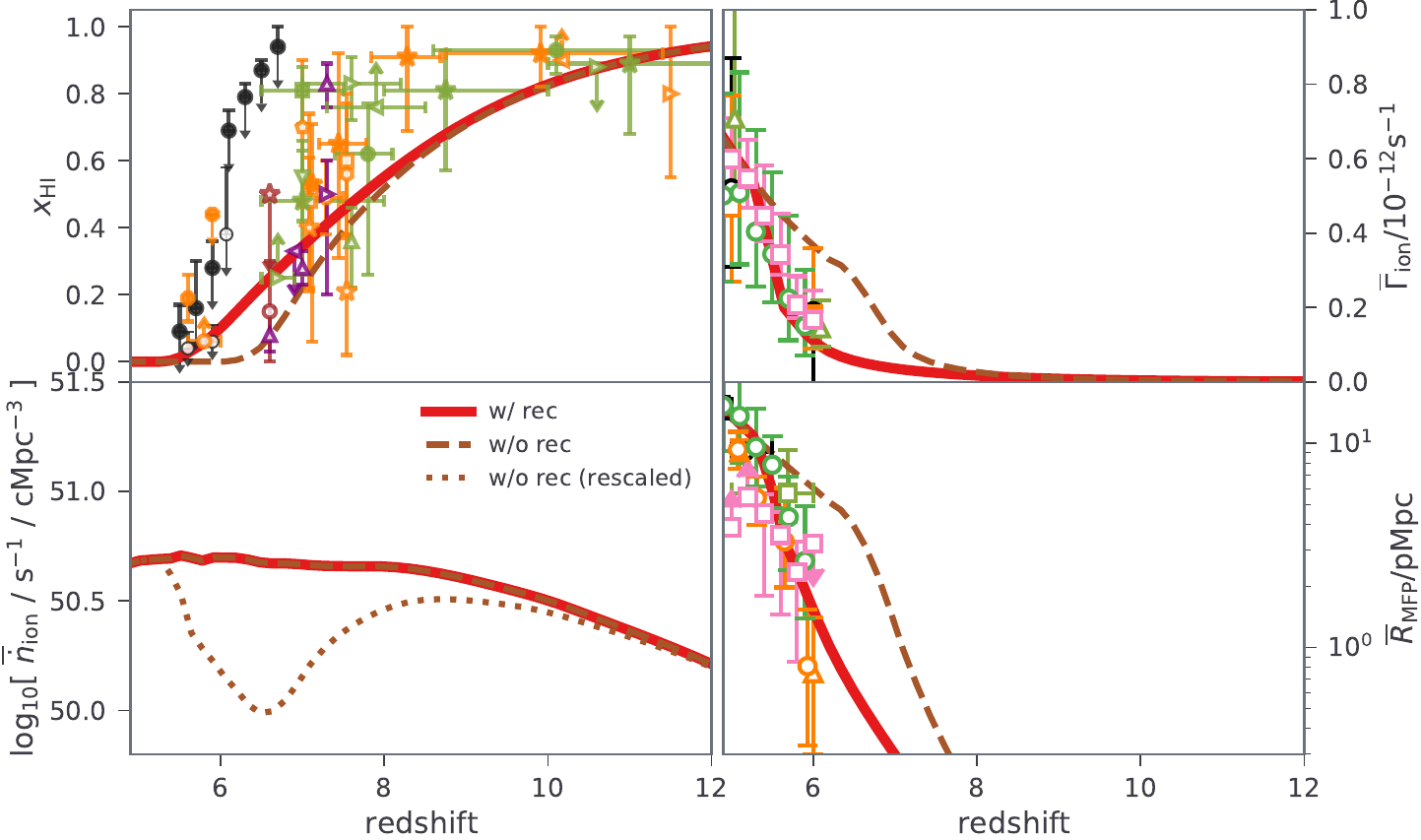}\vspace*{-2.3mm}
		\caption{{\color{black}Comparison of the MAP model with and without recombinations. Clockwise from the upper left panel, we show the mean EoR history, photoionization rate, proper MFP and ionizing emissivity. In the bottom left panel we also show the emissivity rescaled by the ratio of $\overline{\Gamma}_{\rm ion}$ from {\it w/ rec} to that from {\it w/o rec}, roughly mimicking what would be required for the emissivity to compensate for the missing recombinations.}
  \label{fig:rec}
  		}
	\end{minipage}
\end{figure*}

{\color{black} To better quantify this claim, we rerun the MAP model,  turning off inhomogeneous recombinations. Fig. \ref{fig:rec} shows the predicted mean EoR history, photoionization rate, MFP and ionizing emissivity. In the absence of sub-grid recombinations, the end of reionization progresses significantly more rapidly, leading to a correspondingly sharp rise in both the MFP and photo-ionization rate (see also \citealt{Sobacchi2014MNRAS.440.1662S}). Since our models fix the post-reionization MFP to $R_{\rm MFP, LLS}$, these quantities eventually asymptote to the same values as in the fiducial run. In the emissivity sub-panel, we also adjust the emissivity by rescaling it with the ratio of $\overline{\Gamma}_{\rm ion}$ from {\it w/ rec} to that from {\it w/o rec}. We see that by matching the UVB (which roughly corresponds to what forest observations would require for the {\it w/o rec} case), the emissivity would need to decrease rapidly during the second half of the EoR, countering the premature rise in the MFP caused by the lack of recombinations. This lends further credibility to our claim that unresolved, inhomogeneous recombinations are responsible (at least in part) for the rapid drop in the emissivity required by some large-scale hydro simulations in order to match the forest data.}

We note from Fig. \ref{fig:post_emissivity} that our inferred emissivity is consistent with simple estimates assuming a constant escape fraction and integrating the observed UV LFs down to $M_{\rm UV} = -13$.  This is somewhat coincidental, since in our fiducial model more than half of the ionizing photons are provided by galaxies fainter than $M_{\rm UV} > -13$ due to a strong $M_{\rm UV}$-dependence of the ionizing escape fraction (see later Fig. \ref{fig:post_fionfesc}).  As discussed further in Section \ref{sec:model_dependence}, the forest data combined with UV LFs strongly constrain the redshift evolution of the EoR and the ionizing emissivity.  However, determining which galaxies produce the ionizing photons responsible is more model dependent.

\subsection{Effective clumping factor in HII regions}

Modelling the complex interplay between ionizing sinks and sources during the EoR is best achieved with large-scale numerical simulations.  However, simple analytic estimates of the EoR history can be very convenient and help build physical intuition.  A common choice is the following (e.g., \citealt{Bouwens2015b}):
\begin{equation}\label{eq:clumping}
    \dot{Q}_{\rm HII} = \dot{n}_{\rm ion,H} - Q_{\rm HII}  / t_{\rm rec, H}
\end{equation}
where
\begin{equation}
    \overline{t}_{\rm rec, H} {\equiv}  \frac{1}{C_{\rm eff} \overline{n}_{\rm H}\alpha_{\rm B}}
    {=} \left(\frac{1{+}z}{6}\right)^{-3}\left(\frac{C_{\rm eff}}{3}\right)^{-1}\left(\frac{T}{10^4{\rm K}}\right)^{0.75}{\rm Gyr} .
\end{equation}
Here, $Q_{\rm HII} {\sim} 1-\overline{x}_{\rm HI}$ and $\dot{Q}_{\rm HII}$ are the volume filling factors of HII regions and its growth rate while $t_{\rm rec, H}$ is a characteristic recombination time-scale parameterized by an ``effective'' clumping factor, $C_{\rm eff}$.  The first term on the right-hand side of equation (\ref{eq:clumping}) is the ionizing emissivity per hydrogen atom while the second term approximates the global recombination rate per hydrogen atom.  This equation is especially useful in high-redshift galaxy studies, as it allows us to connect the ionizing emissivity from galaxies to the EoR history simply by assuming some value of $C_{\rm eff}$ to capture the impact of recombinations.  Common choices for $C_{\rm eff}$  range from $\sim$ 1 -- 10 (e.g., \citealt{Bouwens2015b,Mason2018ApJ...856....2M,Davies2019MNRAS.489..977D,Bruton2023ApJ...949L..40B}).

\begin{figure}[!ht]
	\begin{minipage}{\columnwidth}
		\centering
		\includegraphics[width=0.98\textwidth]{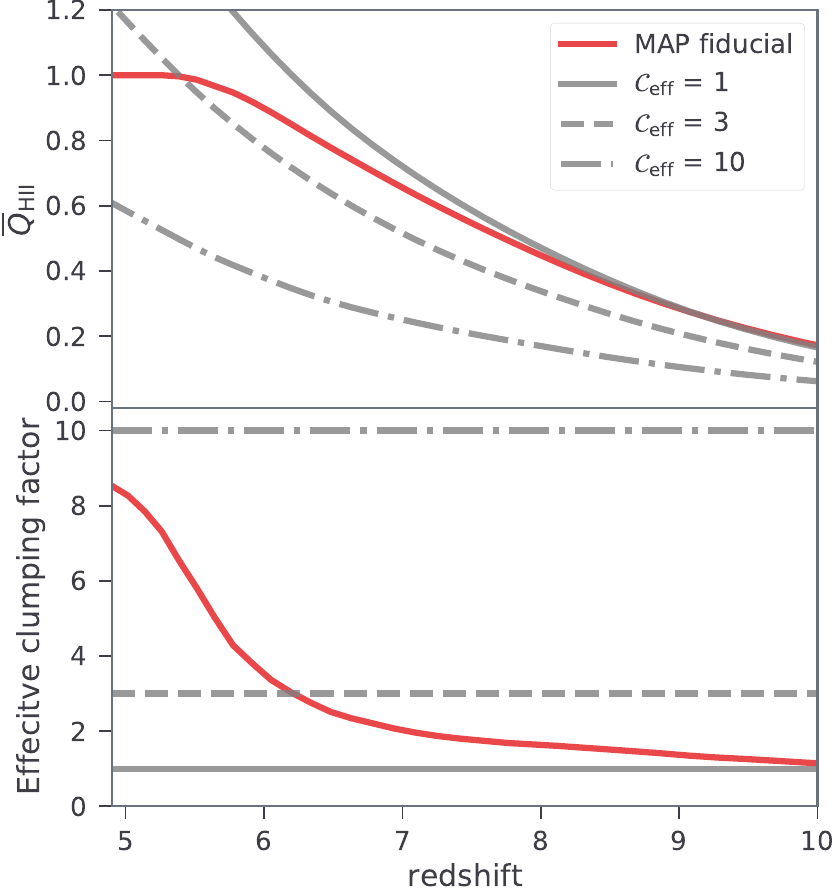}\vspace*{-1.3mm}
		\caption{{\it Top panel:} Evolution of the HII filling factor from the MAP model ({\it red curve}), together with analytic estimates using equation (\ref{eq:clumping}) assuming the same mean ionizing emissivity as the MAP but taking a constant ``clumping factor''. Curves corresponding to $C_{\rm eff}$=1, 3, 10 are shown in grey. {\it Bottom panel:} the effective clumping factor obtained by solving equation (\ref{eq:clumping}) for $C_{\rm eff}(z)$ when assuming the EoR history and emissivity from the MAP model. 
  \label{fig:clumping}
  		}
    
	\end{minipage}
\end{figure}

One can compute the recombination rate in a given patch of the IGM by defining $C_{\rm eff} = \langle n_{\rm HII}^2 \rangle / \langle n_{\rm HII} \rangle^2$, where the averaging is performed over the ionized hydrogen (accounting for self-shielding and using {\it local} values of temperature and ionization rates; e.g. \citealt{Finlator2012MNRAS.427.2464F}).  However, when estimating the {\it global} recombination rate to be used in equation (\ref{eq:clumping}) there is not an obvious way of defining $C_{\rm eff}$ in terms of other global IGM quantities.  In particular, ionizing sources and sinks are strongly correlated on large scales.  Recombinations preferentially occur in regions proximate to galaxies that were the first to reionize, which have biased, time-evolving, and spatially fluctuating properties.

Here, we investigate what choice of $C_{\rm eff}$ can give the same EoR history as the MAP parameter set in our fiducial model.  Specifically, we assume the EoR history and emissivity of our MAP model (c.f. Figures \ref{fig:post_xH} \& \ref{fig:post_emissivity}), and solve for $C_{\rm eff}(z)$ using equation (\ref{eq:clumping}).  The resulting effective clumping factor is plotted as a red curve in the bottom panel of Fig. \ref{fig:clumping}. We see $C_{\rm eff}$ starts around unity\footnote{The clumping factor can also decrease rapidly during earlier EoR stages as the gas relaxes from an increase in the Jeans mass after it is photo-heated (e.g., \citealt{Emberson2013ApJ...763..146E,Park2016ApJ...831...86P,DAloisio2020ApJ...898..149D}). 
However, it is likely that X-ray preheating (e.g., \citealt{HERA2022ApJ...924...51A,HERA2023ApJ...945..124H}) diminishes this evolution in practice.} and then rises rapidly towards the late stages of reionization when ionization fronts penetrate deeper into overdensities (e.g., \citealt{Furlanetto2006PhR...433..181F,Finlator2012MNRAS.427.2464F, Sobacchi2014MNRAS.440.1662S,Cain2024arXiv240902989C,Davies2024arXiv240618186D}).

Fundamentally, $C_{\rm eff}$ cannot be a constant during the EoR.  We illustrate  EoR histories resulting from common assumptions of a constant $C_{\rm eff}$ = 1, 3, 10 in the top panel of Fig. \ref{fig:clumping}.  All curves assume the same emissivity as the MAP.  However, no constant choice of $C_{\rm eff}$ can reproduce the EoR history of the MAP ({\it red curve}). 

We offer a fit for $C_{\rm eff}(z)$ using a rational function (equation \ref{eq:rational}) with coefficients  $\{m_0,m_1,m_2,m_3,n_0,n_1,n_2,n_3\}$ = \{238.9, -94.35, 11.76, -0.404, 22.6, -3.97, -0.877, 0.1636\}.
This can be used in analytic models to approximate the EoR history resulting from a given emissivity. In future work we will quantify how sensitive this effective clumping factor is to different reionization or emissivity models.

\subsection{Galaxy UV LFs and scaling relations}

\begin{figure*}[!ht]
	\begin{minipage}{\textwidth}
		\centering

	\includegraphics[width=\textwidth]{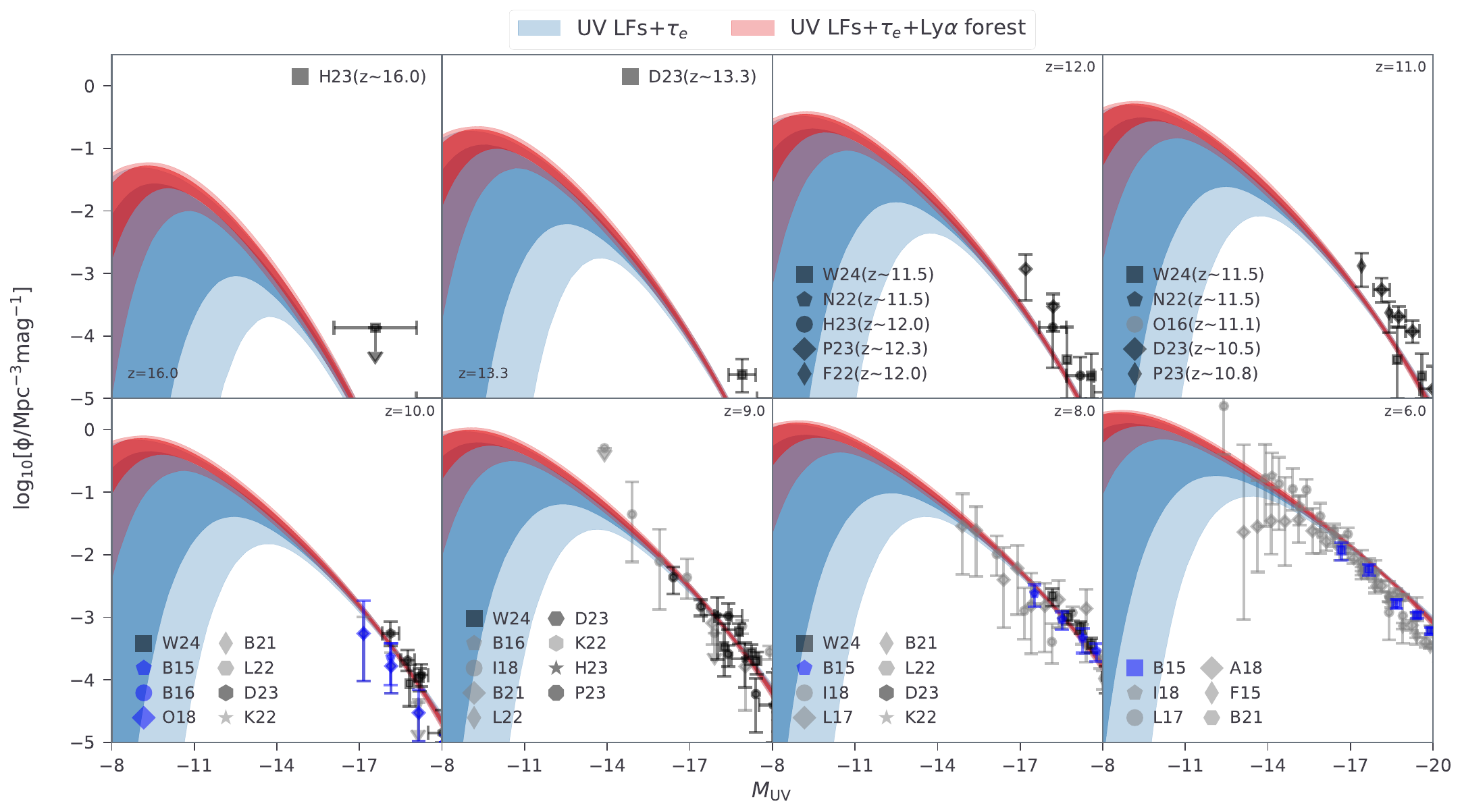}\vspace*{-3.5mm}
		\caption{The inferred galaxy UV luminosity function. As in the previous figure, the dark (light) shaded region corresponds to 68\% (95\%) C.I. Observed luminosity functions are grouped into pre-JWST (light grey; \citealt{Bouwens2015ApJ...803...34B,Bouwens2016ApJ...830...67B,Finkelstein2015ApJ...810...71F,Oesch2016ApJ...819..129O,Oesch2018ApJ...855..105O,Livermore2017ApJ...835..113L,Atek2018MNRAS.479.5184A,Ishigaki2018ApJ...854...73I,Bhatawdekar2019MNRAS.486.3805B,Bouwens2021AJ....162...47B,Kauffmann2022arXiv220711740K,Leethochawalit2022arXiv220515388L}) with those used in the likelihood (see Sec. \ref{subsec:likelihood_other}) highlighted in dark blue, and results using recent JWST observations (dark grey; \citealt{Donnan2023MNRAS.518.6011D,Finkelstein2022ApJ...940L..55F,Harikane2023ApJS..265....5H,Naidu2022ApJ...940L..14N,PG2023arXiv230202429P,Willott2024ApJ...966...74W}).
         \label{fig:post_gal}
	}
  
	\end{minipage}
\end{figure*}

In Fig. \ref{fig:post_gal} we show the inferred UV LFs for our fiducial model.  As in Fig. \ref{fig:post_xH}, the blue shaded regions correspond to our posterior without including forest data (i.e. only including UV LF data and $\tau_e$), while the red regions additionally include the $\tau_{\rm eff}$ distributions from XQR-30+.  In the figure we also show observational estimates from both {\it Hubble} and {\it JWST}, with blue points highlighting those {\it Hubble} datasets that are used in the likelihood (see section \ref{subsec:likelihood_other}). The MAP model and [16, 84]th percentiles are also listed in Table \ref{tab:lfs}.

From the figure we see that the {\it Hubble} estimates we use in the likelihood already constrain the inferred UV LFs at magnitudes brighter than -17, where we have observational estimates. {\color{black} Overall, the predictions also remain consistent with recent JWST measurements (e.g., \citealt{Donnan2023MNRAS.518.6011D, Harikane2023ApJS..265....5H,Willott2024ApJ...966...74W}), though observations at $10<z\lesssim13$ appear slightly higher.
At $z\sim16$, however, UV variability \citep{Shen2023MNRAS.525.3254S,Nicolic2024arXiv240615237N} sourced by enhanced star formation (e.g., \citealt{Qin2023MNRAS.526.1324Q, Wang2023ApJ...954L..48W, Chakraborty2024JCAP...07..078C}) or differences in stellar populations (e.g., \citealt{Ventura2024MNRAS.529..628V, Yung2024MNRAS.527.5929Y}) may be indeed necessary to explain the observed trends. 
This offset suggests that a redshift evolution in $f_*$ (or $\tau_*$) might also be needed in our model, similar to the adjustments made for $f_{\rm esc}$ (see equation \ref{eq:fesc}). We will explore this further as JWST data continue to mature. On the other hand}, the posteriors in blue widen greatly at fainter magnitudes, since there is no observational consensus regarding a faint-end turn-over (see also \citealt{Gillet2020MNRAS.491.1980G} and \citealt{Atek2024Natur.626..975A}). Once we include the forest data however, the posteriors shrink significantly at the faint end of the UV LF.  The forest data imply significant star formation in galaxies down to the atomic cooling limit ($M_{\rm UV} \sim$ -10).

 \begin{figure}
	\begin{minipage}{\columnwidth}
		\centering
		\includegraphics[width=\textwidth]{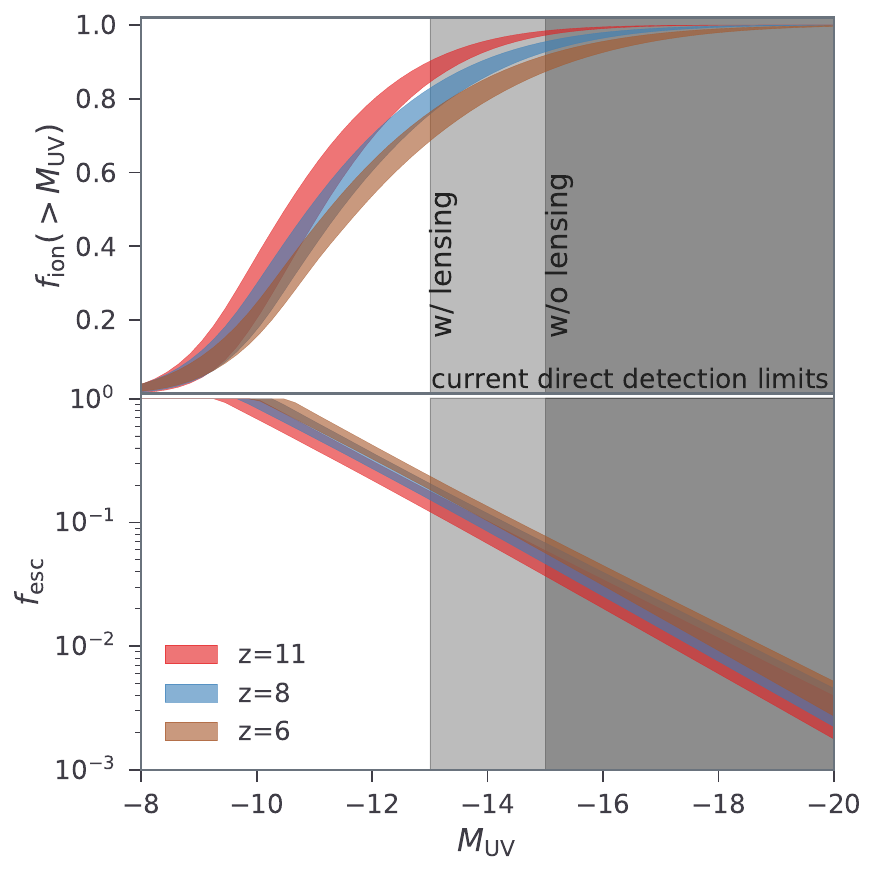}
		\caption{The inferred (68\% C.I.) ionizing contribution of galaxies as a function of their UV mgnitudes at $z=6$, 8 and 11.  The top panel shows {\color{black}the normalized} cumulative number of ionizing photons while the bottom panel shows the escape fraction.  Our results imply reionization is driven by faint galaxies, far below current direct detection limits (roughly corresponding to the grey shaded region). \label{fig:post_fionfesc}
  		}
    
	\end{minipage}
\end{figure}

Note that the Ly$\alpha$ forest is not sensitive enough to directly distinguish between different reionization morphologies (see section 5.3 in \citetalias{Qin2021MNRAS.506.2390Q}).  Therefore the preference for star formation in UV faint galaxies indirectly comes from the {\color{black}mean} EoR history shown in Fig. \ref{fig:post_xH}.  Abundant, faint galaxies appear earlier and evolve more slowly, compared to rare, bright galaxies (e.g., \citealt{Behroozi2019MNRAS.488.3143B}).  Therefore they more naturally drive the kind of slower reionization histories with a ``soft-landing'', which is preferred by the Ly$\alpha$ forest. 
{\color{black} Yet, given sufficient flexibility in assigning ionizing escape fractions, one could in principle force a bright-galaxy-dominated EoR to have the same history as the one shown in Fig. \ref{fig:post_xH}, which is driven by faint galaxies. This, in practice, is constrained by the fact that the escape fraction cannot exceed unity, and that bright galaxies reside in the exponential tail of the mass function.  Thus very extreme evolutions in the ionizing escape fractions, exceeding unity, would be required for our model to have a ``bright galaxy dominated EoR'' that is consistent with Ly$\alpha$ forest data.  Such models are not in our prior volume.
}

We further quantify the contribution of faint galaxies to the EoR in Fig. \ref{fig:post_fionfesc}.  In the top panel we plot the CDF of the galaxies contributing to the ionizing background at $z=$ 6, 8, and 11 as functions of $M_{\rm UV}$.  There is a mild evolution with redshift, but in general we find that galaxies fainter than $M_{\rm UV} \gtrsim -12$ contribute more than half of the ionizing photons that have reionized the universe.  Galaxies above current direct detection limits of $M_{\rm UV} \lesssim -15$ only contribute a few percent to the ionizing photon budget.  This highlights the power of the IGM as a democratic probe of the emissivity of all galaxies.

 In the bottom panel of Fig. \ref{fig:post_fionfesc} we show the {\it mean} ionizing escape fraction as a function of UV magnitude, at the same three redshifts, $z=$ 6, 8, and 11.  We see that the data prefer a population-averaged $f_{\rm esc}$ that increases towards faint galaxies.  This is consistent with most theoretical expectations (e.g., \citealt{Ferrara2013MNRAS.431.2826F,Kimm2014ApJ...788..121K,Paardekooper2015MNRAS.451.2544P,Xu2016ApJ...833...84X}). As the posterior distribution of $\beta_{\rm esc}$ peaks at $\sim-1.6$, the ionizing escape fraction at a given halo mass decreases at earlier times. However, as shown in this panel, such a redshift dependence becomes very mild when $f_{\rm esc}$ is plotted against UV magnitude. 
  The fact that such a mild redshift evolution in the ionizing escape fraction is so strongly preferred by the Bayesian evidence ($\atomicseven$ vs $\atomicsix$) highlights again the incredible constraining power of the XQR-30+ forest data.

\begin{figure*}[ht!]
    \begin{minipage}{\textwidth}
		\centering
	\includegraphics[height=10.7cm]{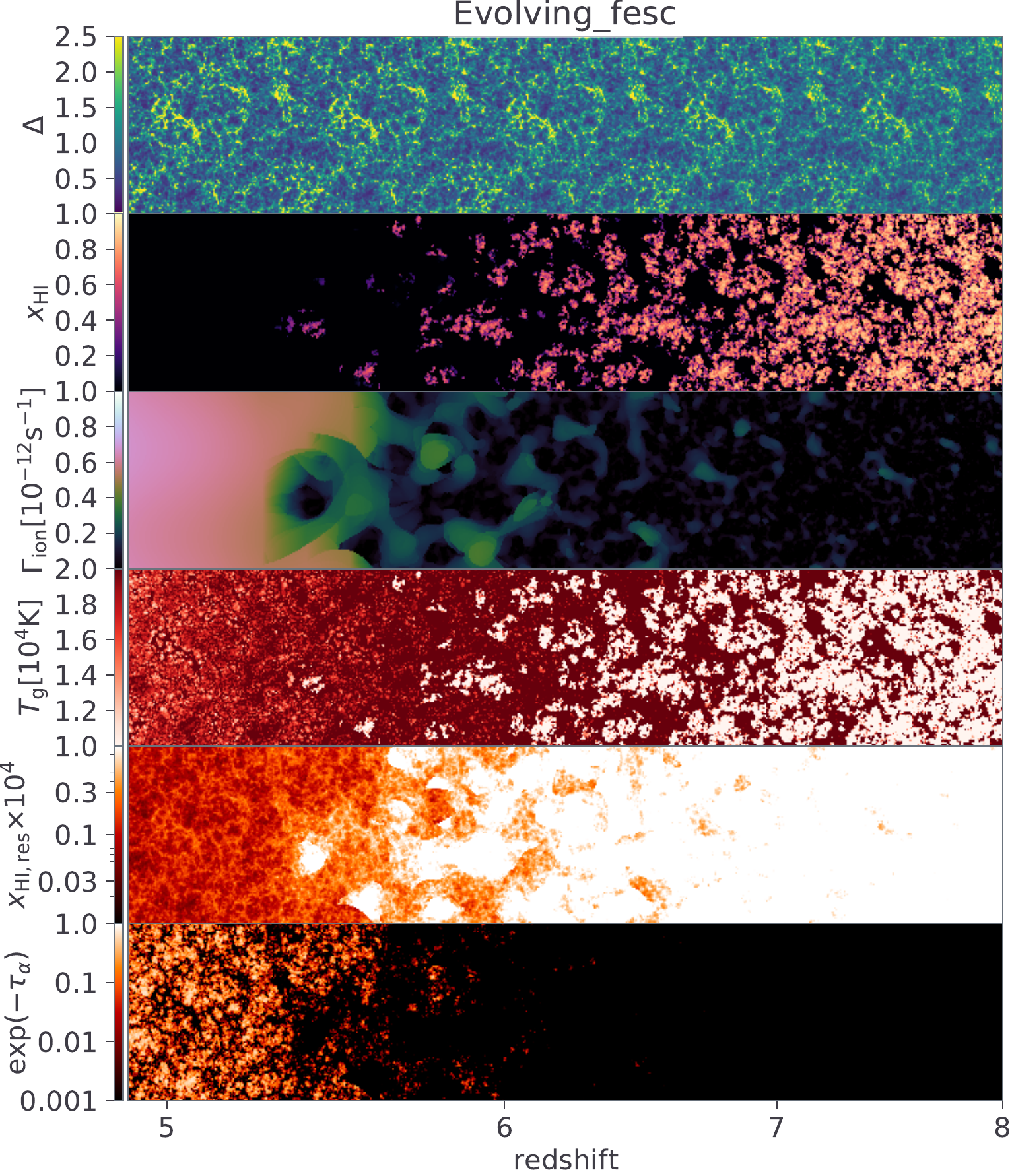}	\includegraphics[height=10.7cm]{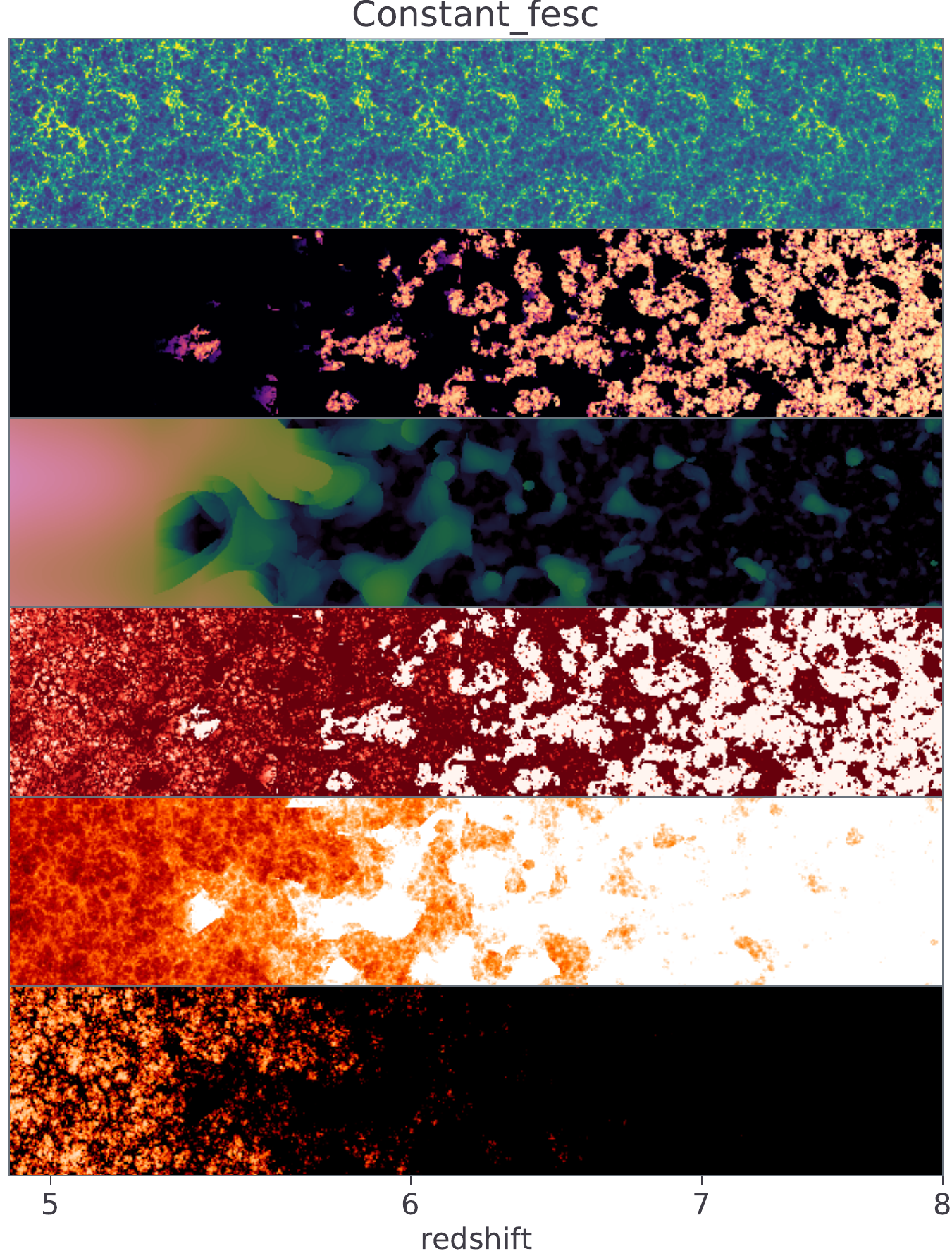}\\
    \includegraphics[width=\textwidth]{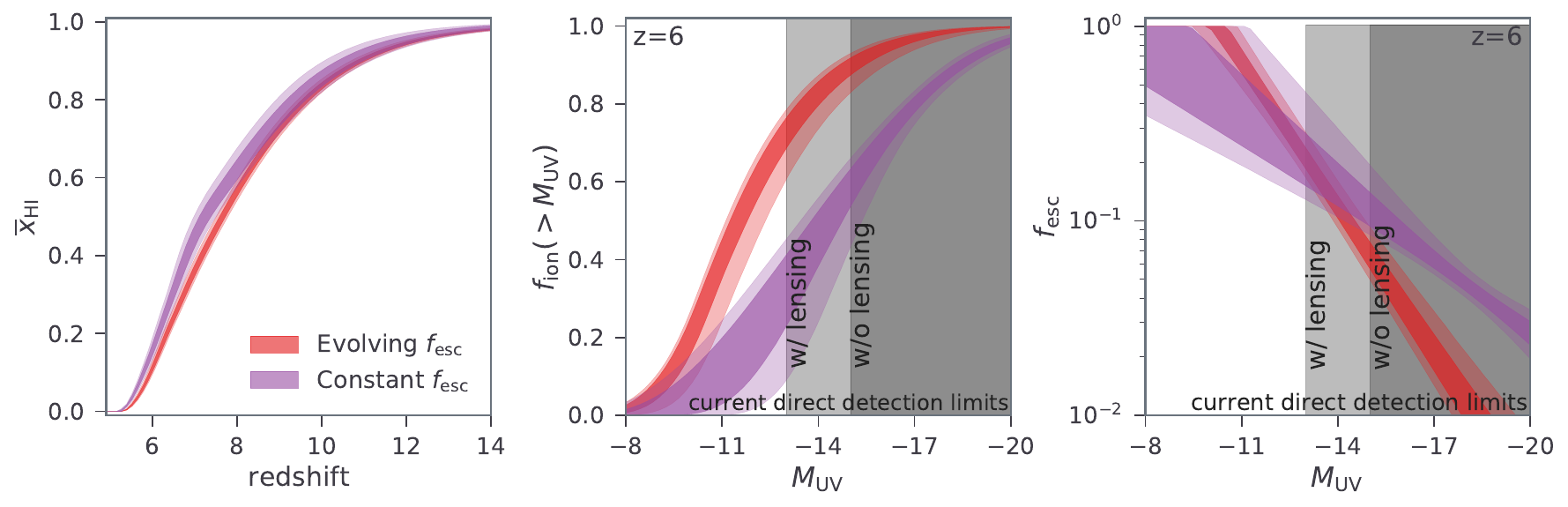}  \vspace*{-5.6mm}
    \caption{{\it Upper panels:} lightcones of MAP models from $\atomicseven$ and $\atomicsix$. From top to bottom, the panels correspond to the overdensity ($\Delta$), neutral hydrogen fraction ($x_{\rm HI}$), locally-averaged UVB ($\Gamma_{\rm ion}$), temperature ($T_{\rm g}$), residual neutral fraction within the ionized regions ($x_{\rm HI,res}$) and Ly$\alpha$ transmission. {\it Bottom panels:} Similar to Figs. \ref{fig:post_xH} and \ref{fig:post_fionfesc} (z=6 only) but for comparisons between $\atomicseven$ and $\atomicsix$ and showing the 68\% and 95\% C.Is of their posterior distributions.
    Although the two models reach qualitatively the same conclusions about the EoR, the fiducial $\atomicseven$ model favors an EoR that is driven by ultra-faint galaxies close to the atomic cooling threshold, resulting in a slightly more extended and patchy EoR.  \label{fig:LCs}}
	\end{minipage}
\end{figure*}

\section{How do the results depend on our model?}
\label{sec:model_dependence}

In the previous section we presented constraints on IGM and galaxy properties using the $\atomicseven$ model, which was strongly favored by the data.
Here, we explore how our main conclusions are affected by the choice of galaxy model. Specifically, we compare our fiducial model to $\atomicsix$ where the UV ionizing escape fraction is solely dependent on the host halo mass.  If our results remain largely unaffected by the choice of galaxy model, this would increase confidence in their robustness, regardless of their relative Bayesian evidences.

Fig. \ref{fig:LCs} shows lightcones corresponding to the MAP in both models ({\it upper panels}) and their posteriors for the EoR history ({\it bottom left panel}), the cumulative contribution to the $z=6$ ionizing background of galaxies below a given UV magnitude ({\it bottom middle panel}), and the ionizing escape fraction as a function of UV magnitude ({\it bottom right panel}). {\color{black} Note that the inferred $\tau_{\rm eff}$ CDFs look indistinguishable to those shown in Fig. \ref{fig:post_CDF}.} Both models suggest a qualitatively similar conclusion -- reionization finishes at $z<5.5$ with the process primarily driven by ionizing photons emitted by faint galaxies. The end (corresponding to $\overline{x}_{\rm HI} = 0.01$) and midpoint of reionization are at $z=5.33\pm0.03$ and $z=7.2\pm0.1$ in the $\atomicsix$ model, respectively, compared to $z=5.44\pm0.02$ and $z=7.7\pm0.1$ in our fiducial $\atomicseven$ model.

From the bottom panels of Fig. \ref{fig:LCs} we see that the models differ quantitatively in which galaxies drive reionization.  The $\atomicsix$ model prefers the EoR to be driven by slightly brighter galaxies, with half of the ionizing photons being contributed by $M_{\rm uv} \gtrsim -14$ galaxies (compared to $M_{\rm uv} \gtrsim -12$ in the fiducial model).  This is due to the fact that without the additional flexibility of a time-evolving $f_{\rm esc}$, the $\atomicsix$ model results in an EoR history that is too rapid compared with what the data prefer.  It is a testament to the constraining power of the XQR-30+ data that only a small redshift evolution in the ionizing escape fraction (c.f. bottom panel of Fig. \ref{fig:post_fionfesc}) results in a much higher Bayesian evidence for the $\atomicseven$ model.

\section{Conclusions}\label{sec:conclusion}

The Ly$\alpha$ forests observed in the spectra of high-redshift quasars provide critical insight into the final stages of reionization. In this work, we introduced a novel framework of {\cmfast} that integrates large-scale lightcones of IGM properties and incorporates unresolved sub-grid physics in the Ly$\alpha$ opacity, calibrated against high-resolution hydrodynamic simulations for missing physics on small scales. By sampling only 7 free parameters that are capable of characterizing the average stellar-to-halo mass relation, UV ionizing escape fraction, duty cycle and timescale of stellar buildup for high-redshift galaxies, we performed Bayesian inference against the latest Ly$\alpha$ forest measurement from XQR-30+ complemented by the observed high-redshift galaxy UV LFs and the CMB optical depth. We demonstrated that current data can constrain global IGM properties with percent-level precision.

One of the key outcomes of our model is the ability to reproduce the large-scale fluctuations in Ly$\alpha$ opacity without requiring a sharp decline in the ionizing emissivity from $z\sim7$ to 5.5, a feature that has been invoked by several other models. In particular, our fiducial model finds reionization occurs at $z=5.44\pm0.02$ with a midpoint at $z=7.7\pm0.1$. The ionizing escape fraction in this model increases towards fainter galaxies, exhibiting only a mild redshift evolution at a fixed UV magnitude. This suggests that half of the ionizing photons responsible for reionization are sourced by galaxies fainter than $M_{\rm UV}\sim-12$, which lie below the detection threshold of current optical and near-infrared instruments including {\it JWST}.

Additionally, we explored an alternative galaxy model that limits the redshift evolution in the ionizing escape fraction, allowing it to only vary with the host halo mass and reducing the number of free parameters to 6. Although this model demonstrates lower Bayesian evidence relative to our fiducial case, the posteriors for the evolution of IGM properties are in qualitative agreement.  This lends confidence that our conclusions on the progress of the EoR are robust.  The models do differ somewhat on which galaxies were driving reionization, with the lower evidence model suggesting galaxies fainter than $M_{\rm UV} \sim -14$ provided half the ionizing photon budget (compared to $M_{\rm UV} \sim -12$ for the fiducial model).

Future observations both in the Ly$\alpha$ forest and in direct galaxy surveys, will be crucial to further refining these models and improving our understanding of which galaxies drive reionization as well as the early stages of the EoR (where we currently only have indirect constraints). Our Bayesian framework, allowing us to connect galaxy properties to IGM evolution in a physically-intuitive manner, represents a significant step forward, offering a versatile and efficient tool for interpreting upcoming observational data.

\section*{Acknowledgement}
The authors gratefully acknowledge the HPC RIVR consortium and EuroHPC JU for funding this research by providing computing resources of the HPC system Vega at the Institute of Information Science (project EHPC-REG-2022R02-213). This work also made use of OzSTAR and Gadi in Australia. YQ acknowledges HPC resources from the ASTAC Large Programs, the RCS NCI Access scheme. YQ is supported by the ARC Discovery Early Career Researcher Award (DECRA) through fellowship \#DE240101129. Parts of this research were supported by the Australian Research Council Centre of Excellence for All Sky Astrophysics in 3 Dimensions (ASTRO 3D), through project no. CE170100013.
A.M. acknowledges support from the Ministry of Universities and Research (MUR) through the PRIN project ``Optimal inference from radio images of the epoch of reionization'', the PNRR project ``Centro Nazionale di Ricerca in High Performance Computing, Big Data e Quantum Computing'', and the PRO3 Scuole Programme `DS4ASTRO'. 
VD acknowledges financial support from the Bando Ricerca Fondamentale INAF 2022 Large Grant “XQR-30”. 
MGH has been supported by STFC consolidated grants ST/N000927/1 and ST/S000623/1.

\section*{Data Availability Statement}
The data underlying this article will be shared on reasonable request to the corresponding author.

\printendnotes
\setlength{\bibitemsep}{1.9pt} 

\printbibliography

\appendix

\section{Detailed posteriors}\label{app:paramter_posterior}

\begin{table*}
\centering
\caption{The inferred neutral fraction, photoionzing rate and MFP for the MAP model and the [16, 84]th percentiles (see also Figures \ref{fig:post_xH} and \ref{fig:post_global}).
 }\label{tab:xhi_gamma_mfp}
\begin{minipage}{0.8\textwidth}  
\centering
  \begin{threeparttable}
\begin{tblr}{
  width = \textwidth,rowsep=0.5pt,
  colspec = {@{} X[c] |  X[c]   X[c]   X[c] |  X[c] |  X[c]   X[c]   X[c] |  X[c] |  X[c]   X[c]   X[c] |  X[c]   X[c]   X[c]   @{} },
  cell{1}{1} = {r=2}{},
  cell{1}{5} = {r=2}{}, 
  cell{1}{9} = {r=2}{}, 
  cell{1}{2} = {c=3}{},
  cell{1}{6} = {c=3}{},
  cell{1}{10} = {c=3}{},
  cell{1}{13} = {c=3}{},
}
\hline
$z$ & $\overline{x}_{\rm HI}$ &&& $z$ & $\overline{x}_{\rm HI}$ &&& $z$ &$\overline{\Gamma}_{\rm ion}/10^{-12}{\rm s}^{-1}$ &&& $\overline{R}_{\rm MFP} / {\rm pMpc}$ \\
\cline{2-4} \cline{6-8} \cline{10-12} \cline{13-15} 
& MAP$^{\dagger}$ & 16th & 84th &    & MAP$^{\dagger}$ & 16th & 84th &    & MAP & 16th & 84th & MAP & 16th & 84th \\
\hline
15.20 & 0.99 & 0.99 & 0.99& 14.88 & 0.99 & 0.99 & 0.99&8.68 & 0.01 & 0.01 & 0.01 & 0.04 & 0.04 & 0.05 \\
14.57 & 0.99 & 0.99 & 0.99& 14.26 & 0.98 & 0.98 & 0.99&8.49 & 0.01 & 0.01 & 0.01 & 0.05 & 0.05 & 0.06 \\
13.96 & 0.98 & 0.98 & 0.98& 13.67 & 0.98 & 0.98 & 0.98&8.30 & 0.01 & 0.01 & 0.02 & 0.06 & 0.06 & 0.08 \\
13.38 & 0.97 & 0.97 & 0.98& 13.10 & 0.97 & 0.97 & 0.97&8.12 & 0.02 & 0.02 & 0.02 & 0.08 & 0.07 & 0.09 \\
12.82 & 0.96 & 0.96 & 0.97& 12.55 & 0.96 & 0.96 & 0.96&7.94 & 0.02 & 0.02 & 0.02 & 0.09 & 0.09 & 0.11 \\
12.29 & 0.95 & 0.95 & 0.96& 12.03 & 0.94 & 0.94 & 0.95&7.77 & 0.02 & 0.02 & 0.02 & 0.12 & 0.11 & 0.14 \\
11.77 & 0.93 & 0.93 & 0.94& 11.52 & 0.92 & 0.92 & 0.93&7.60 & 0.02 & 0.02 & 0.03 & 0.14 & 0.13 & 0.17 \\
11.28 & 0.91 & 0.91 & 0.92& 11.04 & 0.90 & 0.90 & 0.91&7.43 & 0.03 & 0.02 & 0.03 & 0.17 & 0.16 & 0.20 \\
10.80 & 0.89 & 0.89 & 0.90& 10.57 & 0.87 & 0.87 & 0.89&7.26 & 0.03 & 0.03 & 0.03 & 0.21 & 0.20 & 0.25 \\
10.34 & 0.86 & 0.86 & 0.87& 10.12 & 0.84 & 0.84 & 0.85&7.10 & 0.03 & 0.03 & 0.04 & 0.26 & 0.25 & 0.30 \\
 9.90 & 0.82 & 0.82 & 0.84& 9.69  & 0.80 & 0.80 & 0.82&6.94 & 0.04 & 0.04 & 0.04 & 0.32 & 0.30 & 0.37 \\
 9.48 & 0.77 & 0.77 & 0.79& 9.27  & 0.75 & 0.75 & 0.77&6.78 & 0.04 & 0.04 & 0.05 & 0.40 & 0.38 & 0.45 \\
 9.07 & 0.72 & 0.72 & 0.75& 8.87  & 0.70 & 0.70 & 0.72&6.63 & 0.05 & 0.05 & 0.06 & 0.50 & 0.47 & 0.57 \\
 8.68 & 0.67 & 0.67 & 0.69& 8.49  & 0.64 & 0.64 & 0.66&6.48 & 0.06 & 0.06 & 0.07 & 0.62 & 0.59 & 0.71 \\
 8.30 & 0.61 & 0.61 & 0.63& 8.12  & 0.57 & 0.58 & 0.60&6.34 & 0.07 & 0.07 & 0.08 & 0.79 & 0.74 & 0.89 \\
 7.94 & 0.54 & 0.54 & 0.57& 7.77  & 0.51 & 0.51 & 0.54&6.19 & 0.08 & 0.08 & 0.09 & 1.02 & 0.97 & 1.15 \\
 7.60 & 0.47 & 0.48 & 0.50& 7.43  & 0.44 & 0.44 & 0.47&6.05 & 0.10 & 0.10 & 0.11 & 1.37 & 1.29 & 1.53 \\
 7.26 & 0.40 & 0.41 & 0.44& 7.10  & 0.37 & 0.37 & 0.40&5.91 & 0.13 & 0.12 & 0.14 & 1.86 & 1.77 & 2.04 \\
 6.94 & 0.33 & 0.34 & 0.36& 6.78  & 0.29 & 0.30 & 0.33&5.78 & 0.16 & 0.16 & 0.18 & 2.55 & 2.44 & 2.78 \\
 6.63 & 0.26 & 0.26 & 0.29& 6.48  & 0.22 & 0.23 & 0.25&5.64 & 0.20 & 0.21 & 0.23 & 3.32 & 3.46 & 3.89 \\
 6.34 & 0.19 & 0.19 & 0.21& 6.19  & 0.15 & 0.15 & 0.17&5.51 & 0.32 & 0.30 & 0.33 & 5.65 & 5.32 & 5.86 \\
 6.05 & 0.11 & 0.12 & 0.14& 5.91  & 0.08 & 0.08 & 0.10&5.39 & 0.43 & 0.42 & 0.45 & 8.23 & 7.94 & 8.47 \\
 5.78 & 0.05 & 0.05 & 0.07& 5.64  & 0.03 & 0.03 & 0.04&5.26 & 0.54 & 0.53 & 0.57 & 10.91 & 10.88 & 11.06 \\
 5.51 & 0.01 & 0.01 & 0.02& 5.39  & 0.00 & 0.00 & 0.01&5.14 & 0.58 & 0.56 & 0.61 & 12.12 & 12.12 & 12.32 \\
 5.26 & 0.00 & 0.00 & 0.00& 5.14  & 0.00 & 0.00 & 0.00&5.02 & 0.62 & 0.60 & 0.65 & 13.46 & 13.46 & 13.68 \\
 5.02 & 0.00 & 0.00 & 0.00& 4.90  & 0.00 & 0.00 & 0.00&4.90 & 0.66 & 0.64 & 0.70 & 14.96 & 14.95 & 15.19 \\
\hline
\end{tblr}
\begin{tablenotes}

\item[$\dagger$] The inferred MAP $\overline{x}_{\rm HI}(z)$ can be represented by a simple ratio of two polynomials with high accuracy ($|\Delta 
 \overline{x}_{\rm HI}|<0.01$) using the following functional form:
     $x_{\rm HI}(z) = (292.6 - 105.47z + 7.824z^2 +0.312z^3) / (-24.3 + 22.9z-4.96z^2 + 0.694z^3)~$.
\end{tablenotes}
\end{threeparttable}
\end{minipage}

\end{table*}

\begin{table*}
	\centering
	\caption{The inferred galaxy UV luminosity functions for the MAP model and the [16, 84]th percentiles (see also Fig. \ref{fig:post_gal}).
 }\label{tab:lfs}
 \begin{tblr}{
  width = 0.9\textwidth,rowsep=.5pt,
  colspec = {@{} X[c] |  X[c] |  X[c]   X[c]   X[c] |  X[c] |  X[c]   X[c]   X[c] |  X[c] |  X[c]   X[c]   X[c] |  X[c]|  X[c]    X[c]   X[c]   @{} },
  cell{1}{1} = {r=2}{}, 
  cell{1}{2} = {r=2}{}, 
  cell{1}{6} = {r=2}{}, 
  cell{1}{10} = {r=2}{},
  cell{1}{14} = {r=2}{},
  cell{3}{2} = {r=13}{},
  cell{3}{6} = {r=13}{},
  cell{3}{10} = {r=13}{}, 
  cell{3}{14} = {r=13}{}, 
  cell{16}{2} = {r=13}{}, 
  cell{16}{6} = {r=13}{}, 
  cell{16}{10} = {r=13}{},
  cell{16}{14} = {r=13}{},
  cell{1}{3} = {c=3}{},
  cell{1}{7} = {c=3}{},
  cell{1}{11} = {c=3}{},
  cell{1}{15} = {c=3}{},
}
\hline
$M_{\rm UV}$& $z$&$\log_{10}[\phi/\rm Mpc^{-3} mag^{-1}]$&&& $z$& $\log_{10}[\phi/\rm Mpc^{-3} mag^{-1}]$&&& $z$& $\log_{10}[\phi/\rm Mpc^{-3} mag^{-1}]$&&& $z$& $\log_{10}[\phi/\rm Mpc^{-3} mag^{-1}]$\\
\cline{3-5} \cline{7-9} \cline{11-13} \cline{15-17} 
& & MAP & 16th & 84th &  &MAP & 16th & 84th &  &MAP & 16th & 84th &  &MAP & 16th & 84th \\
\hline
-20.0&16&-8.40 &-8.51 &-8.33&13.3&-6.54 &-6.63 &-6.48&12 &-5.75 &-5.83 &-5.70&11 &-5.19 &-5.26 &-5.14\\
-19.0& &-7.18 &-7.27 &-7.15& &-5.57 &-5.65 &-5.54& &-4.89 &-4.96 &-4.87& &-4.41 &-4.47 &-4.39\\
-18.0& &-6.12 &-6.22 &-6.10& &-4.73 &-4.80 &-4.71& &-4.14 &-4.20 &-4.12& &-3.73 &-3.78 &-3.71\\
-17.0& &-5.20 &-5.30 &-5.17& &-3.98 &-4.06 &-3.96& &-3.47 &-3.54 &-3.46& &-3.11 &-3.18 &-3.10\\
-16.0& &-4.38 &-4.50 &-4.35& &-3.31 &-3.41 &-3.29& &-2.87 &-2.96 &-2.85& &-2.56 &-2.64 &-2.55\\
-15.0& &-3.65 &-3.79 &-3.62& &-2.72 &-2.83 &-2.69& &-2.33 &-2.43 &-2.31& &-2.06 &-2.15 &-2.04\\
-14.0& &-3.00 &-3.16 &-2.97& &-2.18 &-2.30 &-2.15& &-1.84 &-1.95 &-1.82& &-1.61 &-1.71 &-1.59\\
-13.0& &-2.43 &-2.60 &-2.41& &-1.70 &-1.84 &-1.68& &-1.40 &-1.53 &-1.38& &-1.20 &-1.31 &-1.18\\
-12.0& &-1.95 &-2.13 &-1.93& &-1.29 &-1.44 &-1.27& &-1.02 &-1.16 &-1.00& &-0.84 &-0.97 &-0.82\\
-11.0& &-1.57 &-1.78 &-1.56& &-0.96 &-1.14 &-0.95& &-0.71 &-0.88 &-0.70& &-0.54 &-0.70 &-0.54\\
-10.0& &-1.34 &-1.63 &-1.33& &-0.75 &-1.00 &-0.75& &-0.52 &-0.74 &-0.51& &-0.35 &-0.56 &-0.35\\
-9.0& &-1.37 &-1.88 &-1.29& &-0.76 &-1.19 &-0.70& &-0.50 &-0.90 &-0.46& &-0.33 &-0.69 &-0.29\\
-8.0& &-1.83 &-2.79 &-1.57& &-1.13 &-1.93 &-0.92& &-0.83 &-1.56 &-0.65& &-0.62 &-1.28 &-0.45\\
\hline
-20.0&10&-4.67 &-4.74 &-4.63&9&-4.20 &-4.26 &-4.16&8&-3.78 &-3.83 &-3.74&6&-3.09 &-3.13 &-3.05\\
-19.0& &-3.97 &-4.02 &-3.95& &-3.57 &-3.62 &-3.55& &-3.21 &-3.26 &-3.19& &-2.64 &-2.67 &-2.62\\
-18.0& &-3.35 &-3.40 &-3.34& &-3.01 &-3.05 &-3.00& &-2.71 &-2.75 &-2.69& &-2.23 &-2.26 &-2.22\\
-17.0& &-2.79 &-2.85 &-2.78& &-2.50 &-2.55 &-2.49& &-2.24 &-2.29 &-2.23& &-1.85 &-1.88 &-1.84\\
-16.0& &-2.28 &-2.35 &-2.27& &-2.03 &-2.10 &-2.02& &-1.82 &-1.87 &-1.81& &-1.49 &-1.53 &-1.48\\
-15.0& &-1.82 &-1.90 &-1.80& &-1.61 &-1.68 &-1.59& &-1.42 &-1.49 &-1.41& &-1.15 &-1.21 &-1.14\\
-14.0& &-1.40 &-1.49 &-1.38& &-1.21 &-1.30 &-1.20& &-1.06 &-1.14 &-1.04& &-0.83 &-0.90 &-0.82\\
-13.0& &-1.01 &-1.12 &-0.99& &-0.85 &-0.95 &-0.84& &-0.72 &-0.81 &-0.70& &-0.54 &-0.61 &-0.52\\
-12.0& &-0.67 &-0.80 &-0.66& &-0.53 &-0.65 &-0.52& &-0.42 &-0.52 &-0.40& &-0.26 &-0.35 &-0.25\\
-11.0& &-0.39 &-0.54 &-0.39& &-0.26 &-0.40 &-0.26& &-0.16 &-0.28 &-0.15& &-0.02 &-0.13 &-0.01\\
-10.0& &-0.21 &-0.40 &-0.20& &-0.08 &-0.26 &-0.07& &0.03 &-0.14 &0.03& &0.17 &0.03 &0.17\\
-9.0& &-0.17 &-0.50 &-0.14& &-0.03 &-0.33 &-0.00& &0.09 &-0.19 &0.11& &0.26 &0.04 &0.27\\
-8.0& &-0.42 &-1.03 &-0.28& &-0.24 &-0.80 &-0.12& &-0.08 &-0.58 &0.03& &0.18 &-0.21 &0.24\\
\hline
\end{tblr}
\end{table*}

\begin{figure*}
    \begin{minipage}{\textwidth}
		\centering
	\includegraphics[width=0.95\textwidth]{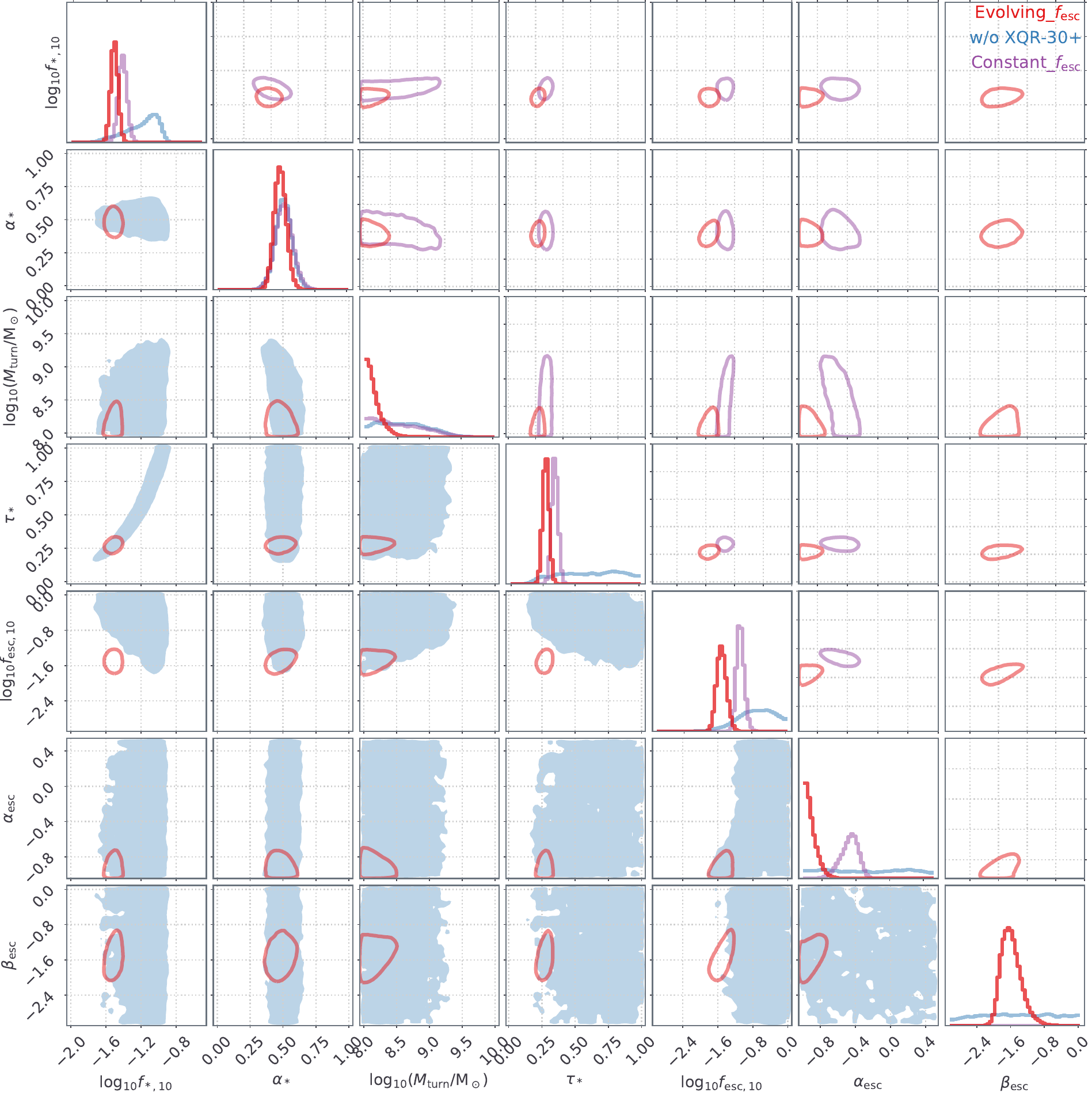}\\\vspace*{-0.8mm}
		\caption{Marginalized 1D and 2D posterior distributions of model parameters from the fiducial model $\atomicseven$ (red), and this model without XQR-30+ (blue) as well as $\atomicsix$ (purple). Regions inside the curves or indicated in shades represent the 95th percentiles.  
         \label{fig:post_paramter}
	}
  
	\end{minipage}
\end{figure*}

Fig. \ref{fig:post_paramter} presents the marginalized 1D and 2D posterior distributions of the model parameters of various models discussed in the work, including the fiducial model $\atomicseven$, and this model without XQR-30+ data, as well as $\atomicsix$. Tables \ref{tab:xhi_gamma_mfp} and \ref{tab:lfs} list the inferred neutral fraction, UVB, MFP and galaxy UV LFs for the MAP model and [16, 84]th percentiles of $\atomicseven$.

\end{document}